\documentclass[runningheads]{llncs}

\usepackage{eccv}

\usepackage{eccvabbrv}
\usepackage{siunitx}
\usepackage{graphicx}
\usepackage{booktabs}
\usepackage{nicefrac}
\usepackage{upgreek}

\usepackage[accsupp]{axessibility}  % Improves PDF readability for those with disabilities.
\DeclareMathOperator*{\argmax}{arg\,max}

\usepackage[dvipsnames]{xcolor}

\usepackage[pagebackref,breaklinks,colorlinks,citecolor=eccvblue]{hyperref}

\usepackage{orcidlink}

\begin{document}

\newcommand{\mytitle}{Single-Photon 3D Imaging with Equi-Depth Photon Histograms}
%\title{3D Vision with Equi-Depth Photon Histograms} 
\title{\mytitle}


\author{Kaustubh Sadekar\orcidlink{0000-0001-8895-5319} \and
David Maier\orcidlink{0009-0002-9146-5411} \and
Atul Ingle\orcidlink{0000-0002-3695-5891}}

\authorrunning{K. Sadekar et al.}

\institute{Portland State University, Portland OR 97201, USA\\
\texttt{\{ksadekar, maier, ingle2\}@pdx.edu}}

\maketitle

\begin{abstract}
Single-photon cameras present a promising avenue for high-resolution 3D imaging.
They have ultra-high sensitivity---down to individual photons---and can record photon arrival times with extremely high (sub-nanosecond) resolution.
Single-photon 3D cameras estimate the round-trip time of a laser pulse by forming equi-width (EW) histograms of detected photon timestamps.
Acquiring and transferring such EW histograms requires high bandwidth and in-pixel memory, making SPCs less attractive in resource-constrained settings such as mobile devices and AR/VR headsets.
In this work we propose a 3D sensing technique based on equi-depth (ED) histograms.
ED histograms compress timestamp data more efficiently than EW histograms, reducing the bandwidth requirement. 
Moreover, to reduce the in-pixel memory requirement, we propose a lightweight algorithm to estimate ED histograms in an online fashion without explicitly storing the photon timestamps.
This algorithm is amenable to future in-pixel implementations.
We propose algorithms that process ED histograms to perform 3D computer-vision tasks of estimating scene distance maps and performing visual odometry under challenging conditions such as high ambient light.
Our work paves the way towards lower bandwidth and reduced in-pixel memory requirements for SPCs, making them attractive for resource-constrained 3D vision applications. Project page: \href{https://www.computational.camera/pedh}{https://www.computational.camera/pedh}
\end{abstract}
    
\section{Introduction}
\label{sec:intro}

The demand for high-resolution, low-cost 3D sensing is growing for a wide range of applications, from autonomous robots to augmented reality, machine vision, surveillance, and industrial inspection.
Thanks to their extreme sensitivity, high spatial resolution (exceeding 100's of kilo-pixels \cite{canonspad,yoshida2021}) and increasing commercial availability, single-photon cameras (SPCs) based on single-photon avalanche diode (SPAD) technology are increasingly popular for dense 3D-sensing LiDARs and on mobile devices \cite{lidar, smartphone}.
SPAD-based SPCs combine the extreme sensitivity of SPADs and the time-of-flight principle to capture scene distance maps with sub-centimeter resolution.

\begin{figure*}[!t]
\centering
\includegraphics[width=0.99\linewidth]{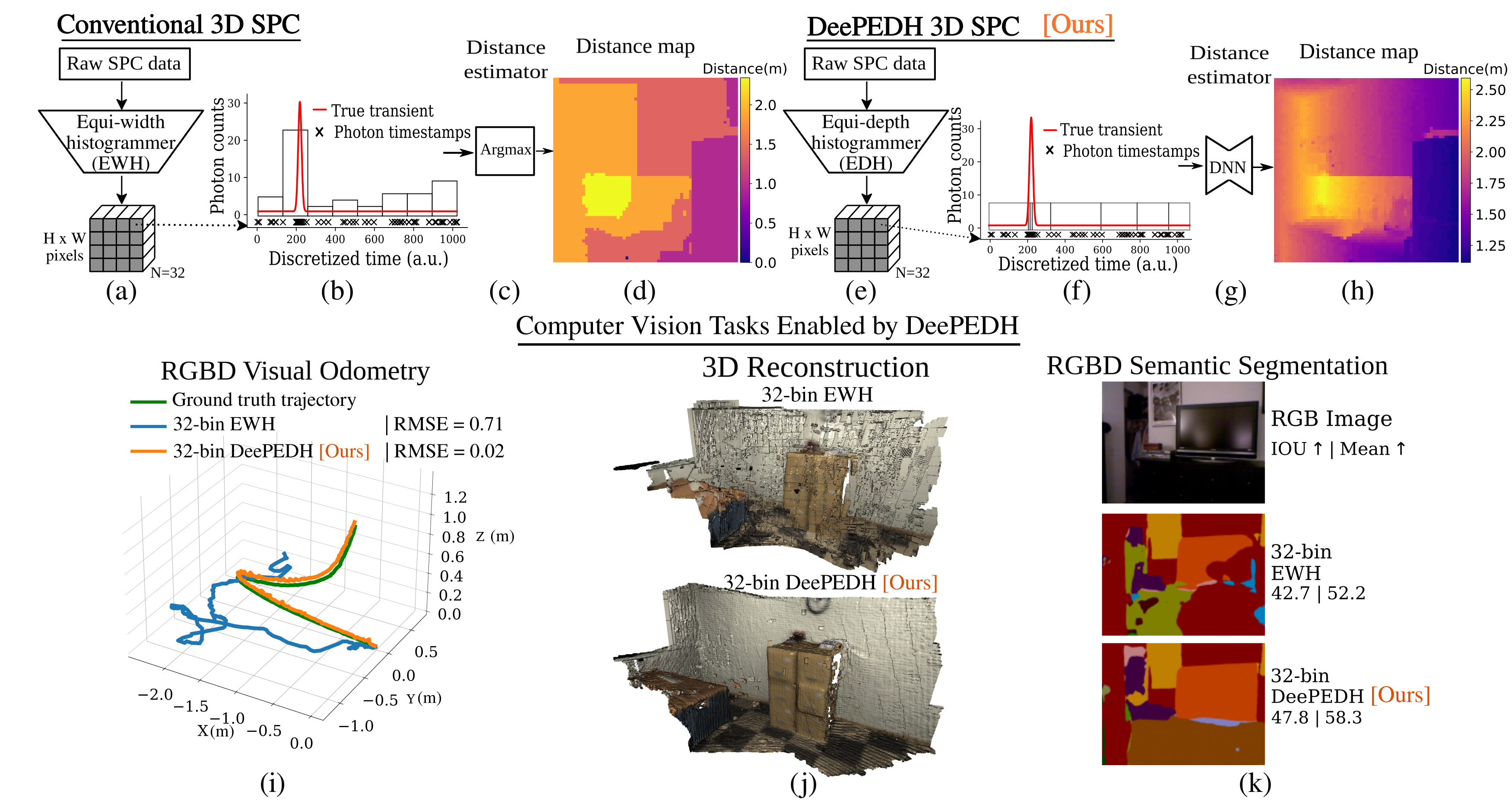}
% \includegraphics[width=0.99\linewidth]{Images/Figure1_draft1_wide_noapp.png}
% \includegraphics[width=0.99\linewidth]{Images/Figure1_draft1.png}
% \includesvg[width=0.8\linewidth]{svg_files/Figure1_draft1.svg}

\caption{\textbf{Our proposed DeePEDH pipeline enables SPCs to be used in resource-constrained applications.}
  (a--c) Conventional 3D sensing pipeline constructs equi-width (EW) histograms that require high in-pixel memory and cause a data bottleneck.
  (d) Existing methods resort to low resolution EW histograms at the cost of poor distance resolution.
  (e--g) DeePEDH uses more efficient on-sensor compression scheme through equi-depth (ED) histograms combined with a deep neural network distance map estimator.
  (h) DeePEDH provides accurate high resolution distance maps with $10-100\times$ lower bandwidth. (i--k) Various downstream vision tasks can benefit from high quality distance maps generated using DeePEDH.}
\label{fig:teaser}
\end{figure*}

Each SPC pixel measures the round-trip time taken for a laser pulse to travel from the camera to the scene and back.
A conventional SPC pixel must repeatedly sample hundreds-to-thousands of photon timestamps to reconstruct the shape of the true laser peak (which we call the \emph{transient distribution}) to estimate the round-trip time-of-flight.
The raw data captured by an SPC can be thought of as a stream of photon timestamps at each pixel location, generating a spatio-temporal ``photon data cube'' \cite{lee2023caspi} shown in Fig.~\ref{fig:teaser}(a).
Each SPAD pixel can detect millions of photon timestamps per second.
Due to hardware constraints, it is impossible to practically process this raw photon data in the sensor or transfer it to any off-sensor compute module.

%As a SPAD-based \textit{single-photon camera} (SPC) can measure accurate photon arrival times, in theory, measuring the round-trip time is as simple as illuminating the scene with a light pulse and capturing the photon-arrival timestamp with the SPC. 
%However, in practice, the SPC also receives photons from ambient light sources.
%Moreover, the light pulse consists of more than one signal photon, distributed around the peak of the signal.
%The peak location is not known a priori, as it depends on the scene distance, which makes it challenging to determine if a specific timestamp corresponds to a signal or a background photon.
%Therefore, a conventional SPC needs to store all the photon timestamps it detects over the exposure time, and estimate the peak of the measured timestamp distribution, to compute the round-trip time.
%Since the camera captures a stream of photon timestamps at each pixel location, the raw data captured by the SPC can be thought of as a photon ``data cube''.
%As SPADs are highly sensitive and extremely fast, they detect millions of photon timestamps per second.
%Hence, the overall photon data cube becomes practically impossible to handle in real-time. 
%
% Hence the data measured by an SPC can be represented as a $H \times W \times B$ data cube where $H \times W$ is the spatial resolution and $B$ represents the number of photon-arrival timestamps. As SPADs are highly sensitive and extremely fast, they detect millions of photon timestamps per second. Hence, the overall data cube of raw photon timestamps becomes practically impossible to handle in real-time. 

%TODO[Done] increase DPI of this image; it looks like 150ish dpi. I use at least 300, or even 500 dpi. 

Conventional SPC pixels summarize the timestamp data by creating equi-width (EW) histograms that keep track of the number of photons received over equally spaced time bins, spanning the entire distance range.
The peak of this EW histogram serves as an estimate of the time-of-flight and hence the scene distance.
In practice, these EW histograms require $\sim 1000$ bins per pixel.
For recent SPAD-based SPCs that have approached megapixel resolutions, the amount of EW histogram data adds up to several gigabytes per second when operating at video rates.
For SPCs to become mainstream, especially for resource-constrained applications, such as low-power mobile robots and smartphone cameras, it is desirable to reduce the volume of the photon data cube without compromising the spatial or temporal resolution of the final distance maps.
%Various methods have been proposed in the past to deal with the high amount of raw data generated by SPCs, and most of them aim to compress the timestamp data to reduce the overall size of the photon data cube without compromising the spatial resolution of the camera.%However, the spatial resolution is still limited to sub-megapixels due to practical bandwidth restrictions. 
%Different compression techniques \cite{csph, racingspad, spatiotemporal} aim to further compress the timestamp data.
%Some of these methods are purely theoretical and are yet to be realized on physical hardware \cite{csph} whereas some are iterative and are compatible with the hardware, but require longer convergence time \cite{racingspad}.

%Conventional SPCs compress the raw data using an equi-width histogrammer (EWH) (a). Resource-constrained applications limit SPC output (b) to fewer (N) bins per pixel. As the EW histograms are inefficient representations of SPC data, the distance estimates are heavily quantized (d). We propose DeePEDH SPCs which use a proportional equi-depth histogrammer (e) to compute a more efficient representation of the SPC data (f).

In this work, we propose a hardware-compatible method that can compress the timestamp data to 32 or even fewer bins per pixel as opposed to the 1000's of bins needed in conventional EW histogram-based processing techniques.
Our method can still retain useful information for reconstructing high-fidelity distance estimates.
Our work is inspired by recent work on constructing equi-depth (ED) histograms instead of the conventional EW histograms \cite{racingspad}.
We estimate the bin-boundary locations of these ED histograms in a count-free fashion, without explicitly storing the photon timestamps.
Through publicly available datasets, we demonstrate that these ED bin-boundary representations can support high-quality distance maps for a variety of 3D scenes and under challenging illumination conditions.
Our method achieves significant in-pixel memory reduction and lower bandwidth requirements relative to conventional EW histogram-based SPCs.
Our contributions in this paper are threefold: (i) We propose a hardware-friendly, in-pixel algorithm that estimates ED histogram bin boundaries for arbitrary $q$-quantiles without having to store the entire history of photon timestamps. 
(ii) We design a learning-based distance-estimation technique that leverages spatio-temporal correlations in the ED boundary locations to generate high-quality distance maps.
(iii) We present three case studies---RGBD visual odometry, 3D scene reconstruction, and RGBD semantic segmentation---to demonstrate the value of using ED bin boundaries as a resource-efficient scene representation for various downstream computer vision tasks.

%As distance estimation is also a key element for the latter two applications, we perform a detailed study for distance estimation from ED histograms obtained using our method.

%For the distance estimation task, we study two classical methods and one deep-learning method and evaluate their performance on two different publicly available datasets under different challenging scenes and lighting conditions.
%The qualitative results for all three distance estimators show that our proposed method can encode useful statistics of the photon timestamp measured by the SPC, which can be further used to reconstruct high-fidelity distance maps.

%Our work aims at making SPCs mainstream for 3D computer-vision applications in resource-constrained conditions and thus benefiting such applications by providing accurate dense distance estimates whilst maintaining lower power and memory requirements.

\section{Related Work}
\label{sec:relwork}

\textbf{3D imaging with SPCs:} 
Most of the current methods using SPCs for 3D imaging explicitly construct an EW histogram based on photon-arrival timestamps \cite{dtofsp3d}, and use the constructed EW histogram for estimating scene distance \cite{spc3dnaturevk, Rapp21, photonflooded, Gupta2019AsynchronousS3}.
Such EWH-based SPCs face a trade-off between the accuracy of distance estimates per pixel and the number of SPC pixels, due to practical bandwidth limits.
To improve spatial resolution without compromising on accuracy, researchers have proposed more efficient measurement and compression techniques inspired by concepts such as Fourier-domain compression \cite{csph}, compressed-sensing with random projections \cite{compresseddepth}, estimating low-dimensional parametric models \cite{parametric}, or sketching \cite{sketching} to track different statistics of the SPC data. 
In contrast, our approach captures ED histogram bin boundaries using a lightweight online algorithm that does not store the full history of photon timestamps, making it amenable to future in-pixel implementations.

\smallskip
\noindent\textbf{Efficient SPC hardware designs:} Improvements in SPC hardware designs have also been proposed considering the practical limitations.
While some propose to measure the EW histogram in a coarse-to-fine scheme by iteratively zooming into the region-of-interest \cite{zooming1, zooming2, zooming3, adaptivegating}, others have proposed designs that can share the resources on the sensor \cite{resourceshare}. 
Although not discussed in this work, our ED histogram-based method can further benefit from recent hardware-friendly designs that use adaptive zooming to reject background light.
While most of the above-mentioned SPC designs require time-to-digital converters (TDCs) to measure the photon-arrival timestamps, new TDC-less designs have been proposed that process a stream of photons using in-pixel neural networks \cite{tdcless1, tdcless2}. 
Tontini et al.\cite{histogramless} proposed a novel 3D-imaging method using SPCs that does not require constructing a histogram of timestamps.
However, it fails in the presence of strong multipath due to the single-peak assumption.
 
\smallskip
\noindent \textbf{Memory-efficient photon data processing:}
The idea of compressing photon timestamps on the fly using different coding techniques has been shown in simulations to provide $>10\times$ reduction in bandwidth \cite{csph, Tachella2022SketchedRH}.
However, to compress the photon timestamps on the fly, additional information such as a coding matrix \cite{csph} needs to be stored on the imaging sensor such that it can be simultaneously accessed by all the pixels, which is challenging to implement using existing pixel designs.
In contrast, our technique constructs ED histograms of photon timestamps in an online fashion without explicitly storing them.
We use a similar binner element (circuit) as Ingle and Maier \cite{racingspad} to iteratively update the ED histogram, but our optimized update strategy for the binner results in faster convergence, lower uncertainties, and improved distance estimates.
Additionally, we also train a deep neural network (DNN) to improve the distance estimates further for challenging scenarios such as high ambient light which benefits downstream computer vision tasks.

\section{Image Formation Model and Equi-Depth Histogram\label{sec:image_formation}}
We summarize the image-formation model for individual pixels of a SPAD-based 3D camera that we use for generating simulated SPAD measurements from existing RGBD datasets. 
We assume that each SPAD pixel is optically co-aligned and operated in synchronization with a pulsed laser source that emits a periodic train of Gaussian-shaped light pulses.
In addition to these \emph{signal} photons reflected from each scene point, the SPAD pixel also records \emph{background} photons which consist of spurious events due to ambient light (e.g., sunlight) and other sources of electronic noise (e.g., dark counts).

In a low-photon-flux scenario and absence of multipath reflections, the average number of photons measured by each pixel is given by a continuous inhomogeneous Poisson process time-varying intensity $\Phi(t)$ and a period equal to the laser repetition period.
(See supplement.)
We discretize the time axis into $B$ locations.
In the $k^\text{th}$ time interval ($1 \leq k \leq B$),
%\footnote{Assuming we divide the complete timeperiod $T_r$ in $N$ equal intervals then width of each interval is $\Delta t = \frac{T_r}{N}$} is defined as:
$\Phi[k] = \Phi_\text{sig}[k] + \Phi_\text{bkg},$
where $\Phi_\text{sig}[k]$ represents the average number of signal photons for time bin $k$ and $\Phi_\text{bkg}$ represents the average number of background photons per time bin \cite{Lindell1, nonfusiontimeres, photoneff, spadnet, csph}.
We call $\Phi[k]$ the \emph{transient distribution}.
Assuming the peak of the transient distribution is located at a time delay $t$ with respect to the start of the laser cycle, the scene distance is given by $z = ct/2$ where $c$ is the speed of light.
In an ideal scenario of no background light and a narrow laser peak that occupies a single discrete time interval, the time of arrival of a single laser photon would be sufficient to estimate scene distance.
In practice, due the Gaussian peak shape and the presence of sensor noise and ambient light, an SPC pixel must sample the true transient over many laser cycles to estimate distance.

\smallskip
\noindent \textbf{Conventional equi-width (EW) histograms:} 
Most SPAD-based SPC pixels today construct EW histograms of photon counts; the location of the peak of this histogram is used to estimate the scene distance.
The number of photons detected in each EW-histogram time bin is stochastic and is governed by a Poisson process. 
The number of photons in the $k^\text{th}$ histogram bin over $N$ laser cycles is given by $\widehat\Phi[k] \sim \mathrm{Poisson}(N \Phi[n]).$
The volume of this histogram data for a 1 megapixel SPC operating at 30 fps exceeds several gigabytes per second---much higher than can be reasonably handled by today's data-transfer buses.
Today's SPCs resort to using fewer pixels and fewer numbers of EW histogram bins to reduce this bandwidth requirement, which lead to poor distance-map estimates that have low spatial resolution and suffer from quantization artifacts.
%TODO[Done] show a distance map with low resolution artifacts in Fig. 1 and failure to perform a CV task as a result. 

\begin{figure*}[!t]
  \centering
   \includegraphics[width=0.9\linewidth]{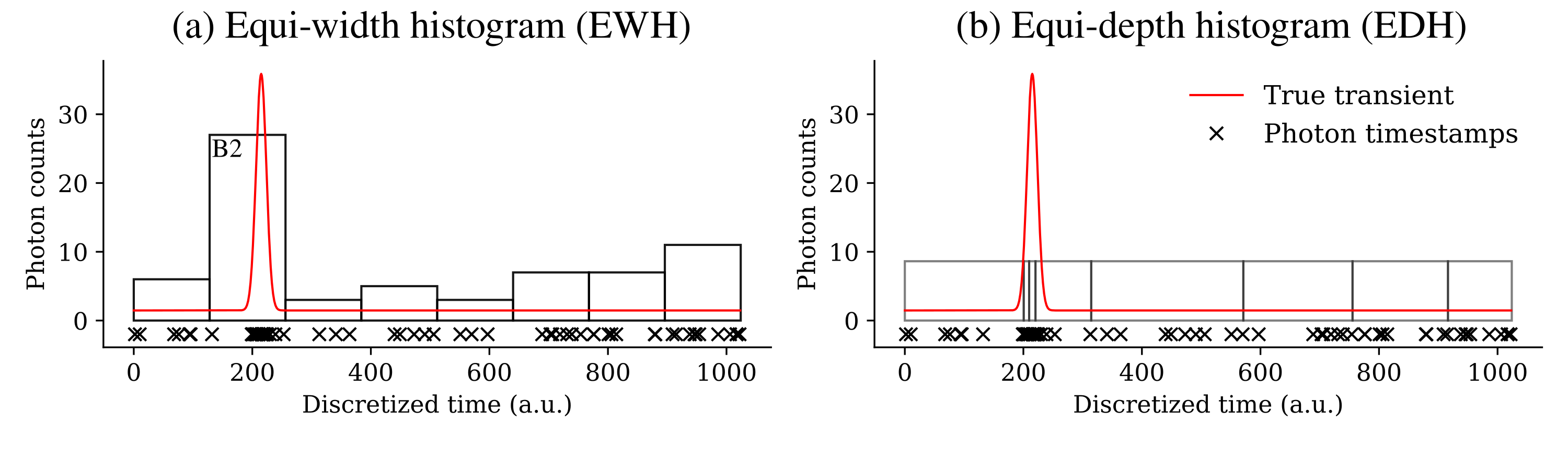}
   \caption{\textbf{Equi-width vs. equi-depth histograms for peaky data.} (a) Most bins of an 8-bin EW histogram are wasted on background photons; a single bin B2 gives a coarse estimate of the true peak.
   (b) An ED histogram captures this ``peaky'' transient distribution better with a cluster of narrower bins around the true peak.}
 %TODO[Done] label EW histogram bin B2 
   \label{fig:ewh-vs-edh}
\end{figure*}

\smallskip
\noindent\textbf{Bandwidth-efficient equi-depth (ED) histograms:} 
Instead of storing photon counts as an EW histogram with equal sized bins, an ED histogram uses variable-width bins such that each bin contains (approximately) equal photon counts.
The term ``depth'' in equi-depth refers to the photon counts in each histogram bin and should not be confused with scene distance.
Fig.~\ref{fig:ewh-vs-edh} shows an example (synthetic) transient distribution consisting of a narrow Gaussian peak.
An 8-bin EW histogram is resource inefficient as seven out of these eight bins are spent on capturing photons from background light.
Moreover, the laser photons all arrive in a single bin $B2$, resulting in an extremely coarse estimate of the scene distance.
In contrast, three of the eight bins in the 8-bin ED histogram cluster around the true peak location.

\begin{figure*}[!t]
  \centering
   \includegraphics[width=0.9\linewidth]{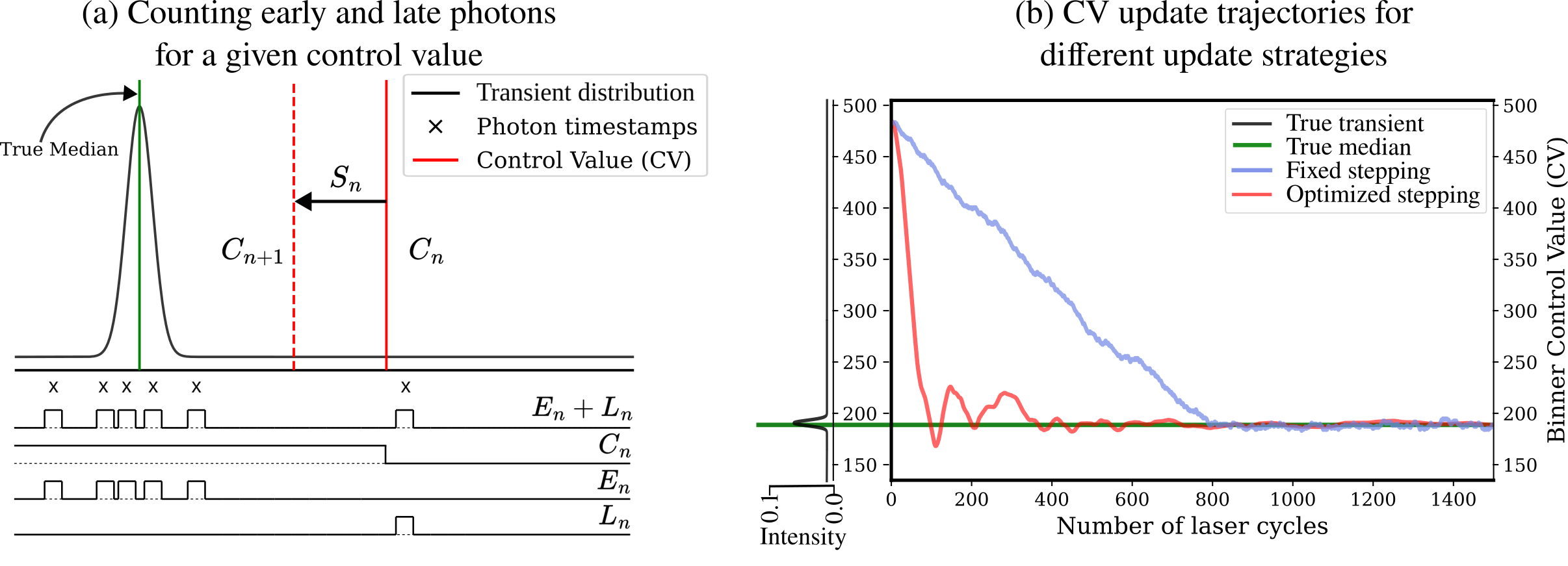}
   \caption{
     \textbf{A binner element used for tracking ED histogram boundaries.}
     (a) A median-finding binner splits incident photons into early ($E_n$) and late photons ($L_n$) compared to the current CV ($C_n$). 
     As the CV is greater than the true median, $E_n/(E_n + L_n) > 1/2$, hence the step size $S_n$ is negative.
     (b) Our optimized stepping strategy uses a sequence of step sizes to achieve faster convergence and lower variance compared to earlier fixed-stepping binner design \cite{racingspad}.}
   \label{fig:papb}
\end{figure*}

Recently ED histograms were proposed as a more resource-efficient alternative to EW histograms for capturing ``peaky'' transient distributions such as that of a laser pulse \cite{racingspad}.
ED-histogram bin boundaries can be estimated using an equi-depth histogrammer (EDH) built using a recursive tree of median-finding \emph{binner circuits}.
A 4-stage tree-based EDH, for example, tracks 15 ED boundaries that constitute 16 quantiles.
A binner circuit maintains an estimate of the median in the form of a \emph{control value (CV)} and iteratively updates it at every laser cycle.
The general update step for the binner in $n^\text{th}$ laser cycle is given by $C_{n+1} = C_n + S_n$ where $C_n$ and $S_n$ denote the CV and the step size for the $n^\text{th}$ laser cycle.
The step size depends on the relative numbers of \emph{early} ($E_n$) and \emph{late} ($L_n$) photons, i.e., photons that arrive earlier or later than the current CV.
The step size is negative when more photons are seen earlier than later, as seen in the example in Fig.~\ref{fig:papb}(a).
% Binner elements based on race logic \cite{racelogic} avoid the need for explicitly digitizing photon timestamps and storing photon counts.
% Unfortunately, this tree-based EDH construction suffers from two drawbacks.
% (i) Binners are constrained to track $q$-quantiles where $q$ is a power of 2.
% (ii) Lower binners in tree-based EDHs receive fewer photons, and their convergence is influenced by the convergence of higher binners.
Ingle and Maier \cite{racingspad} proposed different stepping strategies and the possibility of binner tracking arbitrary $q$-quantiles.
However, several questions remained unexamined in their work and their experimental results were limited to a combination of median-tracking binners that used fixed-stepping. 
We build upon the ideas presented in \cite{racingspad} and propose an optimized binner design called the \emph{proportional binner} that tracks arbitrary $q$-quantiles of the transient distribution with optimized stepping strategies that speed up convergence and reduce variance. 

\smallskip
\noindent\textbf{Proportional binner:}
The stepping strategy of the proportional binner relies on the following key observation: For the CV to track the $j^\text{th}$ quantile (out of $q$), the CV must be updated such that a fraction $j/q$ of photons arrive earlier than the CV and the remaining $1-j/q$ arrive later than the CV.
If the binner had access to the entire history of photon arrival times (say, in the form of a fine-grain EW histogram), the $j^\text{th}$ quantile can be estimated simply by locating the value where the cumulative distribution function crosses $j/q$.
We call a bank of such (hypothetical) binners the \emph{oracle ED histogrammer (OEDH)} and note that in practice, the binner circuit operates in a resource-constrained hardware setting which precludes the possibility of storing the full histogram.

Our proposed stepping strategy for the proportional binner involves iteratively updating its CV at every laser cycle such that the fraction of photons arriving before the CV approaches $j/q$.
At the $n^\text{th}$ laser cycle the control value is updated by a step size proportional to $\Delta_n = \nicefrac{j}{q} - \nicefrac{E_n}{(E_n + L_n)}$ where $E_{n}$ and $L_{n}$ denote the number of photons arriving earlier and later than the current CV.
Although this method does not guarantee deterministic convergence to the $j^\text{th}$ quantile, the stepping strategy is self-correcting --- the larger the magnitude of $\Delta_n$, the larger is the step in the direction that minimizes the discrepency from the $j^\text{th}$ quantile.
%\begin{equation} \label{eq:proportionalstep} 
%\begin{aligned}
%S_n = \left(\frac{K \times N}{100}\right)\Delta_n, \text{ where } \Delta_n = \left(\frac{j}{Q} - \frac{P_E}{P_T}\right).
%\end{aligned}
%\end{equation}
This straightforward stepping strategy of updating CV by $\Delta_n$ suffers from two drawbacks in practice due the presence of strong Poisson noise: (i) 
The CV progresses towards the true quantile location, but continues to wander around that true quantile location.
(ii) Large jumps in the ``wrong'' direction due to background photons cause a large variance in the estimated quantile location. 
We propose an optimized proportional-stepping strategy that mitigates these drawbacks.
First, we apply a \emph{temporal-decay} parameter that gradually reduces the maximum step size and allows the final CV to settle at or close to the true quantile value.
Second, we apply \emph{exponential smoothing} to reduce the variance in the step sizes used for adjusting CV.
The optimized step size is computed using the following equations:
\begin{equation} \label{eq:gammadeltak} 
\begin{aligned}
  S_{n} &= \beta_2 S_{n-1} + (1 - \beta_2) \gamma^n  \widetilde\Delta_{n}\\
  \text{where } \widetilde\Delta_{n} &= \beta_1 \widetilde\Delta_{n-1} + (1 - \beta_1) \Delta_n.
\end{aligned}
\end{equation}
Here $0<\gamma<1$ is the temporal decay parameter, $\widetilde\Delta_n$ is an exponentially smoothed version of the step size $\Delta_n$, and $\beta_1$ and $\beta_2$ are additional tuning parameters that control the level of exponential smoothing applied.
We determine values for these tuning parameters empirically by running extensive simulations of the proportional binner over a wide range of transient distributions.
We find that $\beta_1 = 0.95$, $\beta_2 = 0.8$, and $\gamma = 0.99902$ work well for a wide range of transient distributions over various combinations of signal and background strengths.
(See supplement for details.)
We simulate median-tracking binners with fixed and optimized stepping strategies for a wide range of scene distances and illumination conditions. 
Fig.~\ref{fig:papb}(b) shows CV trajectories for a single run.
Experimental results demonstrate that our optimized stepping binner provides faster convergence over a wide range of illumination conditions than the fixed-stepping binner used by Ingle and Maier \cite{racingspad}. (See supplement.)
% We simulated median tracking binners with fixed and optimized stepping strategies for 10 scene distances and 6 illumination conditions over 500 Monte Carlo runs. 
% As all the binners run for the same exposure of 5000 laser cycles, the lower RMSE between the true median and the binner control values (CVs) shared in Table \ref{table:median-hedhvpedh-t1} demonstrate that our optimized stepping binner converges faster and closer to the true median and is more robust to changing illumination conditions than the fixed-stepping binner used by Ingle and Maier \cite{racingspad}.
%TODO[Done] need to show some example simulation results here in addition to the anecdotal figure? 
% \begin{table}[t]
% \caption{{\small Quantitative metrics for simulated median tracking binners show large improvements in RMSE, faster convergence, and robustness to illumination conditions for optimized stepping.}}
% \vspace{-0.1in}
% \centering
% \setlength{\tabcolsep}{10pt}
% \label{table:median-hedhvpedh-t1}
% {\begin{tabular}{lllll}
% \hline
% Background flux ($\Phi_\text{bkg}$)                                                      & 0.5 & 1.0 & 2.0 & 5.0  \\ \hline
% \begin{tabular}[c]{@{}c@{}}Fixed step \cite{racingspad}\end{tabular}      & 6.50 & 7.32 & 9.39 & 10.13  \\
% \begin{tabular}[c]{@{}c@{}}Optimized step [Ours]\end{tabular}  & 5.46 & 4.20 & 4.95 & 4.06   \\ \hline
% \end{tabular}}
% \vspace{-0.1in}
% \end{table}

A single quantile is often not sufficient to reliably track the peak of a transient distribution. 
We need a bank of binners to track several different quantiles of the underlying distribution. Ingle and Maier \cite{racingspad} use a \emph{hierarchical equi-depth histogrammer (HEDH)}, which consists of a recursive tree of median-tracking binners with fixed-stepping to track different quantiles. However, in HEDH the binners must run sequentially for each level in the tree; convergence of lower-level binners depends on higher-level binners. 

We propose to use a combination of our optimized proportional-stepping binners and configure each one to track a specific quantile of the underlying distribution. We call this design the \emph{proportional equi-depth histogrammer (PEDH)}. The ability of our optimized proportional-stepping binner to track arbitrary quantiles enables the PEDH to run all the binners in parallel and use the complete exposure time. 
Our optimized parameter combination allows the PEDH to track different quantiles with faster convergence and lower variance as compared to HEDH. 
The binner-trajectory plots of 8-bin HEDH and 8-bin PEDH (Fig.~\ref{fig:papb}(b)) illustrate improved accuracy and increased robustness to changing illumination conditions when using PEDH over HEDH. (See supplement for detailed comparison results.)
\section{Distance Estimation from Equi-Depth Histograms}
This section considers the low-level computer-vision task of reconstructing scene-distance maps in a resource-constrained setting.
We first begin with some intuition for how ED-histogram bin boundaries (obtained from a PEDH) can be used to estimate scene distances on a per-pixel basis.
Next, we propose a data-driven approach, called \emph{DeePEDH}, which exploits correlations in ED histogram bins over pixel neighborhoods to further improve the distance-map estimates.

\subsection{Single-Pixel Distance Estimators \label{sec:single_pixel_dist_est}}
The ED histograms bin boundaries inherently capture the density of the underlying transient distribution $\Phi[n]$.
Clusters of narrowly spaced ED histogram bin boundaries correspond to a higher concentration of photon arrivals (such as around the true laser peak location), whereas sparser, widely spaced ED-histogram bin boundaries correspond to a lower concentration of photon arrivals, such as those due to ambient illumination.
Since the ED histogram bin widths are inversely proportional to the local values of the underlying transient distribution, we use the reciprocal of these bin widths as an estimate (up to an unknown scaling factor) of the true underlying transient distribution.
We call these the \emph{local photon density} estimates.
We propose two different distance estimators based on piecewise-constant and piecewise-linear interpolation-based local-photon-density estimates.

For an ED histogram that tracks arbitrary $q$-quantiles, with bin boundary locations given by $\{ t_i \}_{i=0}^q$, we define the piecewise constant local photon density estimator as $\rho_0(t) = \nicefrac{1}{(t_j - t_{j-1})}$ for $t_{j-1} \leq t < t_j$, $1\leq j \leq q$ and $0 \leq t \leq B$.
The extreme ED bin boundaries are, by definition, located at $t_0 = 0$ and $t_q = B$.
Assuming that the transient distribution contains a single sharp peak, we use the midpoint of the narrowest piece of this piecewise constant function as an estimate of the round-trip time-of-flight $t_0$.
Ties are broken arbitrarily.
Let $j^\ast =\argmax _{1\leq j \leq q} \frac{1}{t_j - t_{j-1}}$.
We define the estimate $\widehat t_0 = \nicefrac{(t_{j^\ast} - t_{j^\ast-1})}{2}.$

\begin{figure*}[!t]
\centering
\includegraphics[width=0.9\linewidth]{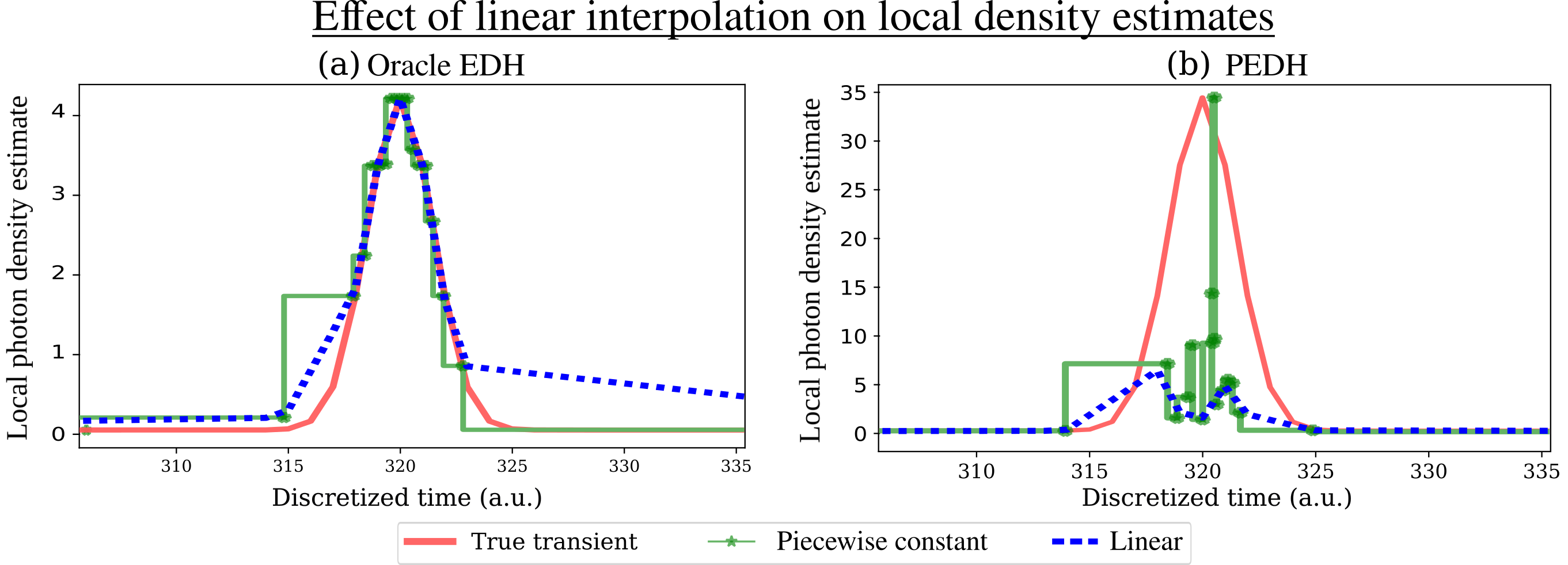}
  \caption{\textbf{Examples of interpolated local density estimates obtained using an oracle EDH and a PEDH.} (a) The narrowest bin and linear interpolation-based estimators perform equally well for this example transient when using the oracle EDH. 
  (b) Linear interpolation-based estimate is often worse than the narrowest-bin estimator due to noise in PEDH bin boundary locations.}
\label{fig:rho0rho1}
%TODO[Done] makes all plot lines thicker 
%TODO[Done] make top figure title with underline: "Effect of interoplation on local density estimates" then make subplot titles "Oracle EDH" and "PEDH".
\end{figure*}

The narrowest-bin-midpoint estimator is computationally lightweight and has the advantage of being amenable to in-pixel implementation.
However, it suffers from a bias away from the true peak location if, for instance, the narrowest and the second narrowest ED bins split the peak, or several narrow bins of equal width cluster around the peak (as seen in the example in Fig.~\ref{fig:rho0rho1}(a)).
Therefore we propose a linearly interpolated local photon density estimator $\rho_1(t)$ which is obtained by taking the sequence of non-uniformly spaced pairs of points $\left\{(\nicefrac{(t_{j}-t_{j-1})}{2}, \nicefrac{1}{(t_j-t_{j-1})} \right\}_{j=1}^q$ interpolated on a grid of $1024$ uniformly spaced discrete time locations between $0$ and $B$.
(See supplement.)
We define an alternative time-of-flight estimate as $\widehat t_1 = \frac{1}{2} \argmax_{t \in [0,B]} \rho_1(t).$
Fig.~\ref{fig:rho0rho1}(a) shows an example of the interpolated photon density estimate $\rho_1(t)$ where the peak lines up with the true peak location, resulting in better distance estimates.

\smallskip
\noindent \textbf{Simulations:}
% To evaluate the performance of the piecewise-constant and piecewise-linear interpolation-based distance estimators, we simulate photon timestamps over $L=5000$ laser cycles for different scenes with varying signal and background levels.
To evaluate the performance of the narrowest-bin-midpoint $\widehat t_0$ estimator and linear interpolation-based $\widehat t_1$ estimator, we simulate photon timestamps over $L=5000$ laser cycles for different scenes with varying signal and background levels.
We use simulated timestamps to estimate ED-histogram bin boundaries using a simulated single-pixel PEDH.
Additionally, we also simulate a single-pixel oracle ED histogrammer (OEDH), which serves as a lower bound for the distance-estimation error achievable with the PEDH.
%TODO[KTBD] check the next two sentences: first we say the two are almost identical, then we're contradicting it by saying that one is better than the other.
We find that the performance of the $\widehat t_0$ estimator is almost identical when using OEDH and PEDH over a wide range of signal and background conditions, which serves as a validation for the optimized PEDH stepping strategies developed in Sec.~\ref{sec:image_formation}.
We observe that the $\widehat t_1$ estimator always performs better than the $\widehat t_0$ estimator with OEDH.
Whereas with the PEDH, $\widehat t_1$ is slightly worse than $\widehat t_0$.
The plots for $\rho_0$ and $\rho_1$ in Fig.~\ref{fig:rho0rho1}(b) provide some intuition.
The converged boundaries for the PEDH are quite close to the OEDH but are not equal due to the inherent randomness of the incident photons and the count-free design of PEDH.
This error in the PEDH boundaries, although small, results in strong noise near the peak of $\rho_0$, which adversely affects the interpolation results, and the peak of $\rho_1$ shifts away from the true peak resulting in sub-optimal distance estimates, increasing the mean absolute error by $\approx$0.1 cm.
See supplement for quantitative accuracy metrics.

\subsection{DeePEDH Distance Estimator}
The per-pixel distance estimators described in the previous section do not exploit correlations across pixel neighborhoods.
Most real scenes have well defined 3D structures, where pixels that are close are likely to have similar distance values.
Moreover, deriving a rule-based method to denoise the PEDH estimates can be difficult, as there is a complex correlation between the scene properties and the sensor properties that determines the variance in the PEDH values.
Hence, we develop a data-driven approach and train a DNN that exploits the spatial (within neighboring pixels) and temporal (within PEDH boundaries of individual pixels) correlations to generate high quality distance maps.

Instead of an end-to-end model (ED histogram boundaries to distance map prediction), we prefer to take a modular approach.
We first compute the per-pixel photon density estimates $\rho_1$ from the measured PEDH boundaries.
%TODO[KTBD] clarify why we use rho_1 here when in the previous paragraph we said that rho_1 is slightly worse than rho_0 for PEDH? 
Next, we train a DNN to denoise $\rho_1$ and predict the scene distance maps.
We call this deep learning-based PEDH-to-distance-map method the \textit{DeePEDH distance estimator}. 
Our approach is inspired by the ``PENonLocal Deep Boosting'' DNN proposed by Peng et al. \cite{photoneff} with one key difference --- instead of training the model on noisy 1024-bin EW histograms, our model is trained on 1024-dimensional linearly interpolated photon density estimates $\rho_1$ obtained from PEDH output.
The loss function used for training the network is a combination of Kullback-Leibler (KL) divergence between the histograms and total variation (TV) distance between the distance maps, identical to the loss function used by Peng et al. \cite{peng2022boosting}. 

% We adapt the recently proposed deep boosting model \cite{photoneff}.
% however the overall idea of using deep-learning to estimate the scene distance from any count-free equi-depth histogrammers has never been studied before. \imp{the above paragraph might be suitable for discussions or related work but I also want to ensure that while reading this section the reader should not lose interest in thinking that we simply retrained a model and claimed it to be a novel idea. NOTE: We are also using the non-local block of the network so the model is almost identical}. The experimental details and evaluation results are presented in the next section.

\section{Experiments}

% \subsection{PEDH Dataset Generation}
\smallskip
\noindent \textbf{PEDH dataset generation.}
We use the image-formation model presented in Sec.~\ref{sec:image_formation} to generate a simulated photon timestamp dataset from existing NYUv2 and Middlebury RGBD datasets \cite{nyuv2, midury}.
We assume a Gaussian laser pulse of \SI{100}{\nano\second} repetition period (corresponding to a maximum distance range of \SI{15}{\m}) and full width at half maximum (FWHM) of \SI{0.32}{\nano\second}.
We simulate the PEDH output with $q=32$ ED bins on a per-pixel basis.
We run each proportional binner for $L = 5000$ laser cycles with the following optimized stepping parameters: $\gamma = 0.99902$, $m_1 = 0.95$, and $m_2 = 0.8$.
(See supplement for details on parameter selection.)
We also generate 1024-bin EW histograms for baseline comparisons.

\begin{figure*}[!t]
\centering
\includegraphics[width=0.95\linewidth]{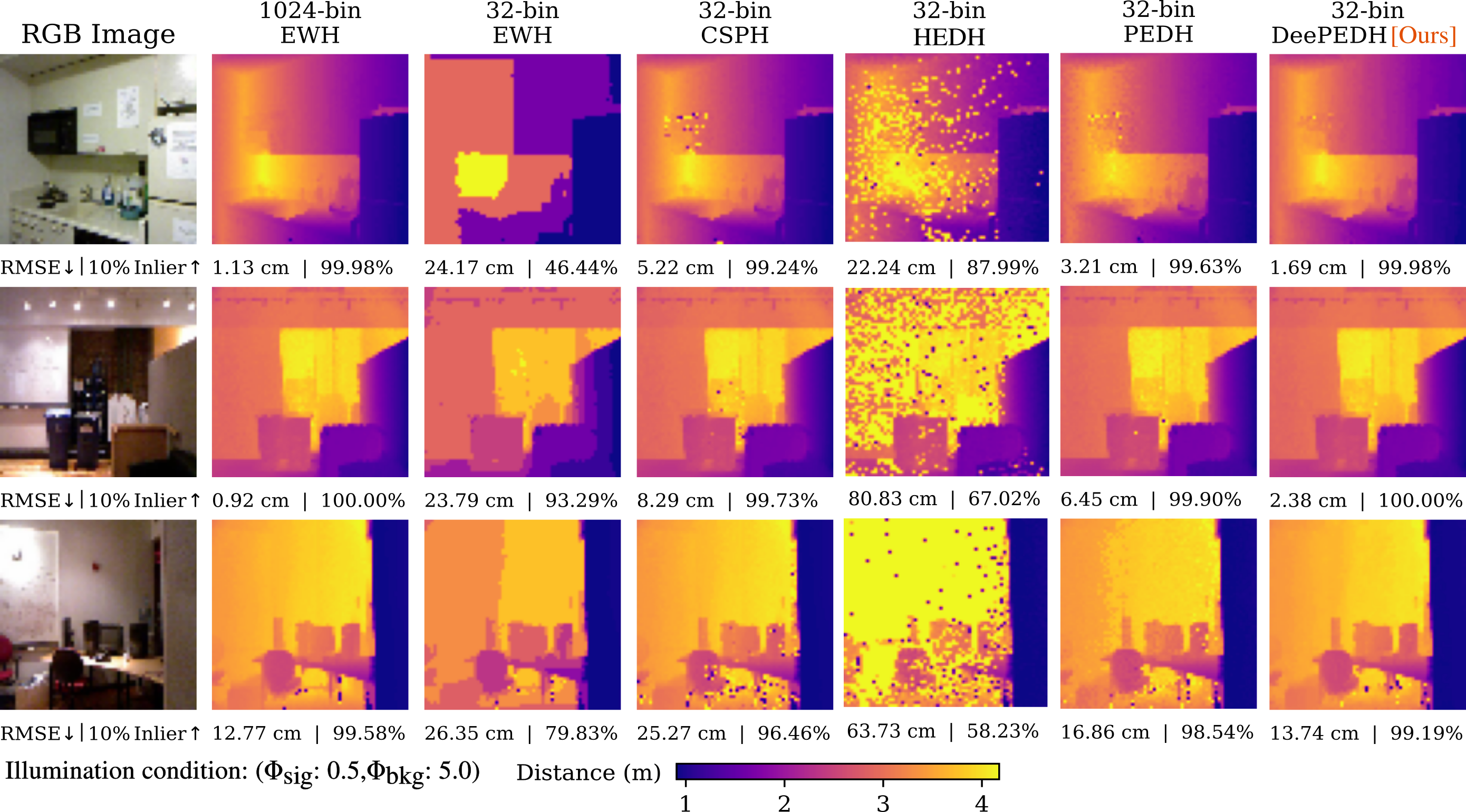}
\caption{\textbf{Comparison of distance map reconstructions on NYUv2 test images.} 
  The CSPH \cite{csph} algorithm, HEDH \cite{racingspad} (narrowest bin), and the PEDH (narrowest bin) methods suffer from noisy distance maps in darker regions and farther distances.
  The 32-bin EWH suffers from quantization artifacts.
  The proposed DeePEDH method provides high spatial-and-distance resolution even in regions where other methods fail.}
  %TODO[Done] Add "RMSE (downarrow) | 10% Inlier (uparrow)" below each RGB image 
\label{fig:distestim_results1}
\end{figure*}
%TODO[Done]: Doublecheck rows 2 and 3 in this figure --- 32 bin EWH results look better than expected? the top row looks more like what I expect with the blotchy appearance. [The images are for 32-EWH, this might be due to the dynamic range. Zooming into the images we can see the blotchy effect]

\begin{table}[t]
\caption{Distance estimation results comparing the average performance of EWH-based (conventional) SPCs, CSPH\cite{csph}, HEDH\cite{racingspad} (narrowest bin), narrowest-bin PEDH distance estimator, and DeePEDH for 10 NYUv2 test set images and 8 different (signal, background) photon-level pairs.}
%TODO[KTBD] I removed the Gbps row as it isn't really a quantitative image quality metric and is based on arbitrary choice of fps and image resolution. 
% \label{quant-table1}
% \begin{tabular}{r|lllll}
%   & \begin{tabular}[c]{@{}l@{}}Conventional\\ (1024-bin)\end{tabular} & \begin{tabular}[c]{@{}l@{}}Conventional\\ (32-bin)\end{tabular} & \begin{tabular}[c]{@{}l@{}}CSPH \cite{csph}\\ (32-bin)\end{tabular} & \begin{tabular}[c]{@{}l@{}}PEDH\\ (32-bin)\end{tabular} & \begin{tabular}[c]{@{}l@{}}DeePEDH \textcolor{RedOrange}{[Ours]}\\ (32-bin)\end{tabular} \\ \hline
% %Bandwidth (Gbps) & 100 & 0.3 & 0.3 & 0.3 & 0.3 \\
% RMSE (cm) & 6.89 & 29.80 & 15.01 & 18.05 & 10.92 \\
% MAE (cm) & 1.06 & 24.58 & 2.03 & 2.40 & 1.84 \\
% 2\% Inliers $\uparrow$ & 99.89 & 10.69 & 99.45 & 97.87 & 98.51 \\
% 10\% Inliers $\uparrow$ & 99.90 & 55.86 & 99.62 & 99.71 & 99.81 \\ \hline
% \end{tabular}
% \end{table}
\label{quant-table1}
\begin{tabular}{r|llllll}
  & \begin{tabular}[c]{@{}l@{}}EWH\\ (1024-bin)\end{tabular} & \begin{tabular}[c]{@{}l@{}}EWH\\ (32-bin)\end{tabular} & \begin{tabular}[c]{@{}l@{}}CSPH \cite{csph}\\ (32-bin)\end{tabular} & \begin{tabular}[c]{@{}l@{}}HEDH \cite{racingspad}\\ (32-bin)\end{tabular} & \begin{tabular}[c]{@{}l@{}}PEDH\\ (32-bin)\end{tabular} & \begin{tabular}[c]{@{}l@{}}DeePEDH \textcolor{RedOrange}{[Ours]}\\ (32-bin)\end{tabular} \\ \hline
%Bandwidth (Gbps) & 100 & 0.3 & 0.3 & 0.3 & 0.3 \\
RMSE (cm) $\downarrow$ & 6.89 & 29.80 & 15.01 & 219.91 & 18.05 & 10.92 \\
MAE (cm) $\downarrow$ & 1.06 & 24.58 & 2.03 & 105.9 & 2.40 & 1.84 \\
2\% Inliers $\uparrow$ & 99.89 & 10.69 & 99.45 & 81.06 & 97.87 & 98.51 \\
10\% Inliers $\uparrow$ & 99.90 & 55.86 & 99.62 & 86.27 & 99.71 & 99.81 \\ \hline
\end{tabular}
\end{table}

% \subsection{Training and Validation}
\smallskip
\noindent \textbf{Training and validation.}
To simulate training and validation datasets, we generate PEDH output for a wide range of scenes and illumination conditions from the NYUv2 dataset \cite{nyuv2}.
The training and validation datasets are generated using the RGBD images from the NYUv2 dataset.
The training dataset consists of 2000 RGBD images from the NYUv2 training set of the dataset and the validation dataset consists of 200 RGBD images from the NYUv2 test set.
To make our DeePEDH model robust to a wide range of signal and background levels, we simulate the PEDH output by randomly choosing a pair of average signal and background photon-levels per cycle from the following pairs: $(\Phi_\text{sig}, \Phi_\text{bkg}) \in \{$(1.0, 1.0), (1.0, 2.0), (1.0, 5.0), (0.5, 0.5), \allowbreak (0.5, 1.0), (0.5, 2.5)$\}.$
Since we do not make any architectural changes to the ``PENonLocal Deep Boosting'' model \cite{photoneff}, the chosen validation dataset is only used to determine training convergence.

% \subsection{Performance Evaluation}
\smallskip
\noindent \textbf{Performance evaluation.}
We evaluate the performance of the narrowest-bin distance estimator $\widehat t_0$ presented in Sec.~\ref{sec:single_pixel_dist_est} and the DeePEDH distance estimator and compare the results with four baseline methods --- peak of the full resolution (conventional) 1024-bin EWH, peak of a coarsely binned  (conventional) 32-bin EW histogram, distance estimated from the compressive single-photon histogram (CSPH) method \cite{csph}, and 32-bin HEDH \cite{racingspad} (narrowest bin).
We evaluate the performance on two different test sets. 
The first test set consists of 10 images from the Middlebury dataset \cite{midury} and the second test set consists of 10 images from the test set of the NYUv2 dataset \cite{nyuv2}.
For each test image, we test the distance estimators for the following eight (signal, background) photon-level pairs:
$(\Phi_\text{sig}, \Phi_\text{bkg}) \in \{$(1.0, 1.0), (1.0, 2.0), (1.0, 5.0), (1.0, 10.0),  (0.5, 0.5), (0.5, 1.0), (0.5, 2.5),  (0.5, 5.0) $\}.$
The performance is compared based on four different evaluation metrics:  RMSE, MAE, $2 \%$ inlier metric, and $10 \%$ inlier metric.
Fig.~\ref{fig:distestim_results1} shows qualitative and quantitative results for three images from NYUv2 test set\cite{nyuv2}. 
Observe the distance images from HEDH \cite{racingspad} are grainy and contain large errors for darker objects at farther distances. The distance images from the narrowest-bin PEDH estimator are less grainy, demonstrating the robustness of PEDH over HEDH to challenging illumination conditions.
DeePEDH reconstructs high quality distance maps even for darker regions in the scenes at farther distances and does not suffer from quantization artifacts.
% Observe the distance images from the narrowest-bin PEDH estimator are grainy, especially for darker objects at farther distances.
Table~\ref{quant-table1} shows quantitative results for  10 images of NYUv2 test set averaged over the 8 different (signal, background) pairs.
These results highlight that the DeePEDH can exploit spatio-temporal correlations present in PEDH boundary estimates, to generalize well on such challenging scenarios and produce high quality distance maps.
Additional results for both NYUv2 and Middlebury datasets are shown in the supplement.

% \subsection{Hardware emulation results}
\smallskip
\noindent \textbf{Hardware emulation results.}
We conducted hardware emulation experiments on a publicly available hardware dataset by Gupta et al. \cite{Gupta2019AsynchronousS3} containing real-world noise sources (dark counts, afterpulsing, dead-time). This data was captured with a single-pixel SPAD LiDAR setup. We emulated (8-bin) HEDH and PEDH for 5000 laser cycles for 5 illumination conditions.
Fig. \ref{fig:hardware_results} shows results for one of the scenes.
(See supplement for additional comparisons.)

\begin{figure*}[!t]
\centering
\includegraphics[width=0.96\linewidth]{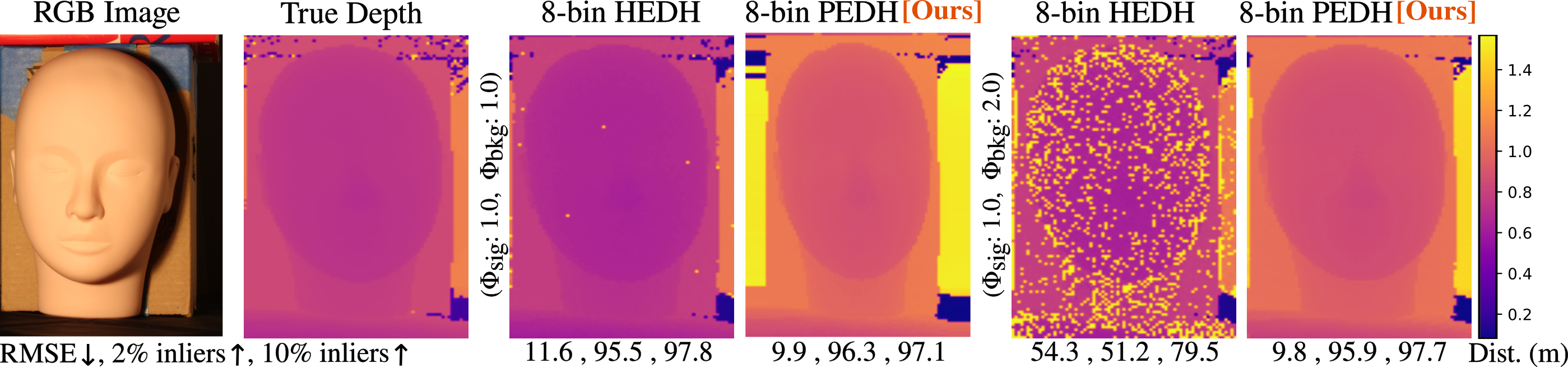}
\caption{\textbf{PEDH improves distance estimates over HEDH.} 
  Experimental results with real-world data show improved distance estimates (RMSE cm,  2\% inliers, 20\% inliers) with our proposed PEDH compared to the HEDH of Ingle \& Maier \cite{racingspad}.}
\label{fig:hardware_results}
\end{figure*}

\section{Equi-Depth Histograms for Computer Vision Tasks}
Do PEDH-based SPCs work well for downstream computer vision tasks in resource-constrained applications?
We show results for three computer vision tasks: RGBD visual odometry (VO), 3D reconstruction using truncated signed distance function (TSDF) fusion, and RGBD semantic segmentation.
These tasks are common components of resource-constrained applications such as autonomous robots, augmented reality applications for mobile devices, and virtual reality headsets.
For all three applications, we simulate distance estimates using 32-bin EWH (conventional SPC) and 32-bin PEDH, so that both use the same amount of in-pixel memory.
The PEDH output is processed using DeePEDH for distance estimation which is then fed to existing processing pipelines. 

\begin{figure*}[!t]
\centering
\includegraphics[width=0.98\linewidth]{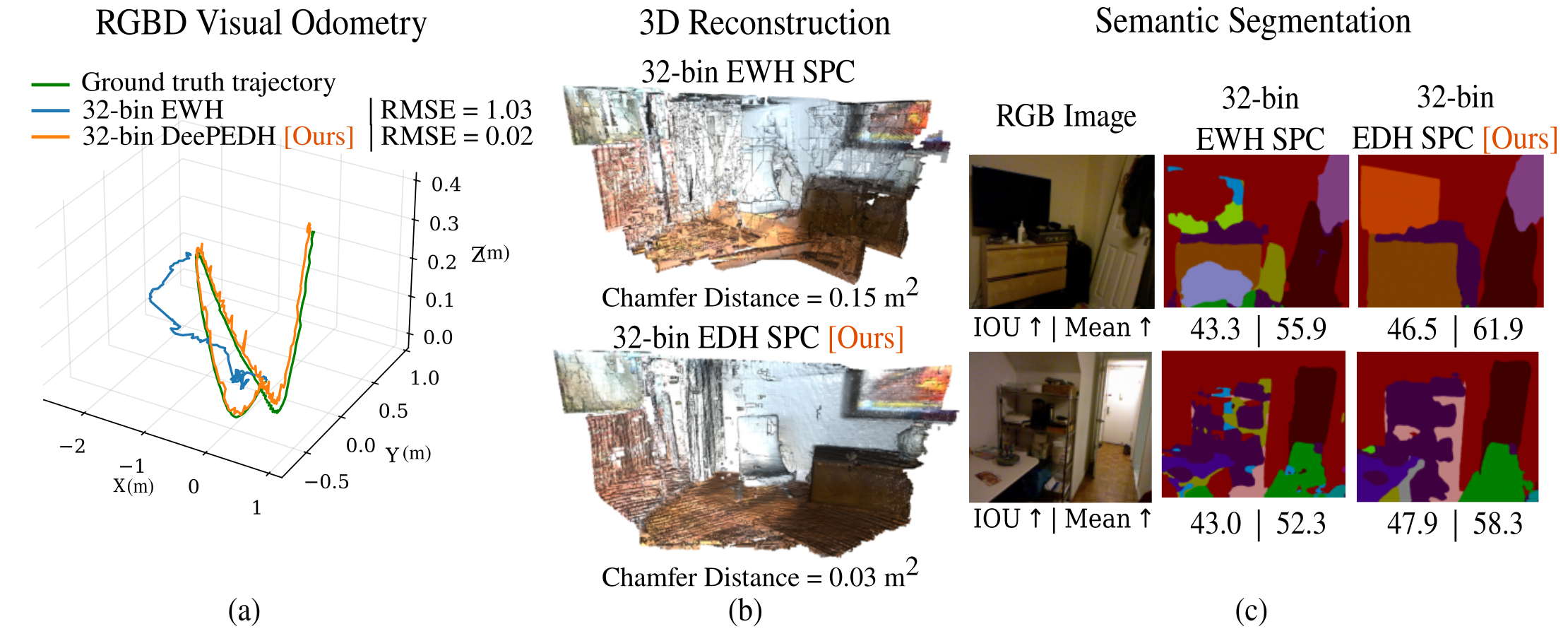}
  \caption{\textbf{DeePEDH SPCs enable better camera tracking, 3D reconstruction, and improve performance of deep RGBD semantic segmentation models.} (a) The camera motion trajectory reconstructed from DeePEDH distance maps (green) closely tracks the ground truth (green) and provides $>10\times$ lower RMSE than 32-bin EWH (blue).
  (b) 3D surface reconstruction obtained using a 32-bin EWH suffers from severe quantization artifacts. The proposed 32-bin DeePEDH method generates high quality 3D reconstructions both qualitatively and in terms of quantitative metrics. 
  (c) semantic segmentation results for a pre-trained CEN network \cite{cen} using distance estimates from 32-bin EWH and 32-bin DeePEDH. 
  As the distance maps from DeePEDH are significantly closer to the ground truth, the segmentation results are better than using distance images from 32-bin EWH.}
\label{fig:voresults}
\end{figure*}

% \subsection{Application 1: RGBD-Visual Odometry}
\smallskip
\noindent \textbf{Application 1: RGBD-visual odometry.}
Visual odometry \cite{odom1, odom2} is the task of estimating camera pose and motion from image sequences captured by the camera.
In the case of RGBD cameras, the sequences consist of RGB images and corresponding distance maps.
The key component of a VO pipeline is the feature-matching step where the movement of interesting visual features is estimated between consecutive frames and used to compute the camera trajectory.
The quality of distance estimates in each frame directly affects the feature-matching step.
For resource-constrained applications there are strict bandwidth and memory limitations that affect the quality of distance measurements and thus the performance of the VO pipeline. We simulate distance maps from 32-bin EWH and 32-bin DeePEDH and pass the RGBD sequences from the Redwood dataset \cite{redwooddataset} to the existing RGBD VO pipeline using the Open3D library \cite{open3d}.
Fig.~\ref{fig:voresults}(a) shows camera trajectories obtained from the VO pipeline using distance maps from both methods.
The 32-bin EWH suffers from large tracking errors due to inaccurate distance estimates that accumulate over the motion trajectory.
In contrast, our method closely tracks the ground truth.

\smallskip
\noindent \textbf{Application 2: Dense 3D reconstruction.}
Dense 3D reconstruction is an important computer vision task for applications such as virtual reality, digital twinning of 3D assets, photo-realistic telepresence, and autonomous mobile robots.
3D reconstruction pipelines have two major components: (i) estimating the camera pose per frame and, (ii) fusing the RGBD information to reconstruct a 3D scene that is consistent with a maximum number of RGBD frames and the corresponding camera poses.
A common method to fuse the RGBD information is the TSDF fusion method \cite{tsdf}. 
We use a similar experimental setup as the VO application and use the distance maps from EWH and DeePEDH as input to the TSDF pipeline of Open3D\cite{open3d}. We pass the ground truth camera poses to the TSDF pipeline instead of the VO estimates.
For quantitative evaluation, the \textit{chamfer distance} is computed between the point cloud reconstructed from each method and the point cloud reconstructed using the ground truth distance maps for TSDF fusion. Fig.~\ref{fig:voresults}(b) shows qualitative and quantitative results for TSDF fusion.
Observe that the EWH results suffer from strong quantization artifacts whereas the DeePEDH-based 3D reconstruction preserves finer details.
% \begin{figure}[htpb]
% \centering
% \includegraphics[width=0.7\linewidth]{Images/Figure10_3DRecon.png}
% \caption{\textbf{PEDH SPCs result in better 3D reconstruction}: Figure showing 3D reconstruction obtained from TSDF fusion using distance estimates from 32-bin EWH SPC (left) and 32-bin PEDH SPC (right). Using a 32-bin PEDH results in cleaner, high-fidelity reconstruction.}
% \label{fig:tsdf-fusion}
% \end{figure}

% \begin{figure*}[!t]
% \centering
% \includegraphics[width=0.72\linewidth]{Images/Figure7.png}
% \vspace{-0.06in}
% \caption{\textbf{PEDH SPCs improve performance of deep RGBD semantic segmentation models}: Figure showing semantic segmentation results for a pre-trained CEN network \cite{cen} using distance estimates from 32-bin EWH and 32-bin DeePEDH.
% As the distance maps from DeePEDH are significantly closer to the ground truth, the segmentation results are better than using distance images from 32-bin EWH.}
% \label{fig:rgbdseg}
% \end{figure*}

% \subsection{Application 3: RGBD deep semantic segmentation}
\smallskip
\noindent \textbf{Application 3: RGBD deep semantic segmentation.}
In semantic segmentation, each pixel of the image is assigned a class label.
Multiple deep-learning methods have been proposed in the past that perform semantic segmentation, some only use RGB image \cite{unet,fcn,deeplab,pspnet,segformer} whereas some use the scene distance map in addition to the RGB image \cite{cen,rgbdjmalik,rgbdrcnn,rgbdmutex}. 
We use the Channel-Exchanging Network (CEN) proposed by Wang et.al.\cite{cen} to compare the effect of using distance maps from the 32-bin EWH SPC and the 32-bin PEDH SPC, on the task of deep RGBD semantic segmentation.
We use the NYUv2 dataset as our test set.
We generate CEN output using the ground-truth distance map, and distance maps obtained from the 32-bin EWH SPC, and DeePEDH.
The CEN output using a ground-truth distance map is used as the ground-truth for the segmentation result to compare the output for the other two cases.
Fig.~\ref{fig:voresults}(c) shows the output for all three cases for different scenes.
We observe that segmentation results are more accurate for DeePEDH as compared to the conventional 32-bin EWH SPC.
% \begin{figure}[htpb]
% \centering
% \includegraphics[width=0.9\linewidth]{Images/Figure13.png}
% \caption{\textbf{PEDH SPCs result in better 3D reconstruction}: Figure showing 3D reconstruction obtained from TSDF fusion using distance estimates from 32-bin EWH SPC (left) and 32-bin PEDH SPC (right). Using a 32-bin PEDH results in cleaner, high-fidelity reconstruction.}
% \label{fig:rgbdseg}
% \end{figure}

\section{Discussion}

%\textbf{Practical circuit design.}
%Although we did not design a hardware binner circuit, the key operations involved in computing the step size at each laser cycle involve only $\sim \!\!3$ multiply-add operations that can be performed cheaply in hardware.
%The initial step-size computation requires a division operation, which could be performed through a look-up table shared across groups of pixels. 
% \noindent\textbf{Future scope for improved EDH.} Our DeePEDH method benefits in three ways over HEDH\cite{racingspad}: optimized stepping in the binner, parallel operation of all the binners, and an advanced DNN for distance estimation. We plan to study and extend these benefits to HEDH. There is also scope to improve the stepping in the binner by combining information from neighboring pixels.

%\noindent\textbf{SPCs for resource-efficient 3D vision.}
In this work, we proposed a resource-efficient method for producing a compact representation for single-photon camera outputs and studied the effect of using these representations for 3D perception in power-and data-constrained applications.
We proposed different distance estimators to predict scene distances from the PEDH output and validate them in both simulation and hardware emulation.
Although we did not design a hardware binner circuit, the key operations involved in computing the step size at each laser cycle involve only $\sim \!\!3$ multiply-add operations that can be performed cheaply in hardware.
The initial step-size computation requires a division operation, which could be performed through a look-up table shared across groups of pixels.
Our experimental results highlight that using our proposed PEDH-based SPCs can achieve superior performance over conventional EWH-based SPCs for popular computer vision tasks like distance estimation, visual odometry, dense 3D reconstruction, and RGBD semantic segmentation.
Our DeePEDH results highlight the power of using deep learning to exploit spatio-temporal correlations captured in PEDH output to improve 3D perception in resource-constrained settings. 

% the results show that deep learning is improved good for 3D perception in  resource constrained settings 

\section*{Acknowledgements}
This work was supported in part by NSF ECCS-2138471.

\bibliographystyle{splncs04}
\bibliography{main}

\clearpage
\renewcommand{\figurename}{Supplementary Figure}
\renewcommand{\tablename}{Supplementary Table}
\captionsetup[figure]{name={Suppl. Fig.}}
\captionsetup[table]{name={Suppl. Table}}
\renewcommand{\thesection}{S.\arabic{section}}
\renewcommand{\theequation}{S\arabic{equation}}
\setcounter{figure}{0}
\setcounter{section}{0}
\setcounter{equation}{0}
\setcounter{table}{0}
\setcounter{page}{1}
\renewcommand*{\thefootnote}{$\dagger$}

\begin{center}
%\begin{tabular}{c}
\huge Supplementary Document for\\
\huge ``\mytitle'' \\[0.7cm]
\large Kaustubh Sadekar, David Maier, Atul Ingle \\[0.7cm]
\texttt{\{ksadekar,maier,ingle2\}@pdx.edu}
\end{center}

\renewcommand*{\thefootnote}{\arabic{footnote}}

\section{Details of the Image-Formation Model \label{suppl:image_formation}}

This section provides details for the image formation model used for generating simulated photon streams for the distance-map estimation methods presented in the main text.
Our simulation model relies on ground truth data from existing RGB-D datasets and simulates photon events using a Poisson process, similar to existing work \cite{Lindell1, nonfusiontimeres, photoneff, spadnet, csph}.

We assume that the SPAD-based 3D camera illuminates each scene point with a pulsed laser that generates a periodic Gaussian pulse train with repetition period $T_r$.
Laser photons reflected from each scene point are captured by a co-aligned SPAD pixel.
The time-varying light intensity $\Phi(t)$ incident on the SPAD pixel is given by:
  \[
    \Phi(t) = g\left( t - \frac{2z}{c}\right) + b_a
  \]
for $t \in [0, T_r)$, where $g(\cdot)$ represents the Gaussian pulse shape, $z$ is the scene distance, $c$ is the speed of light, and $b_a$ is the background (ambient) illumination. 
Since the SPAD pixel has limited time resolution, we discretize the time axis into steps of $\Delta t$, so that there are $B = \frac{T_r}{\Delta t}$ discrete time locations.
In a low-photon-flux scenario and absence of multiple reflections, the mean number of photons received by the SPAD pixel in the $k^\text{th}$ sampling interval of width $\Delta t$ is given by:
\begin{equation} \label{eq:rn} 
\Phi[k] = \eta \int_{k \Delta t}^{(k+1) \Delta t} g\left(t - \frac{2z}{c}\right) \,dt \ + b_{a}\Delta t + b_\text{dark}
\end{equation}
where  $\eta$ captures laser photon losses (due to scene reflectance, distance-squared falloff, cosine-angle falloff, and non-ideal photon detection probability), $b_\text{dark}$ is the number of ``dark counts'' in the time interval $\Delta t$.
We call $\Phi[k]$ the \emph{transient distribution}.
We express Eq.~(\ref{eq:rn}) more compactly as having a time-varying signal term $\Phi_\text{sig}[k]$ and a constant ``background'' term $\Phi_\text{bkg}$, where it is understood that the background term includes sensor dark counts:
\begin{equation} \label{phibar}
\Phi[k] = \Phi_\text{sig}[k] + \Phi_\text{bkg}.
\end{equation}
  The \textit{signal-to-background ratio} (SBR) is defined as $\mathsf{SBR} \overset{\operatorname{def}}{=} \nicefrac{\Phi_\text{sig}} {B\Phi_\text{bkg}}$\footnote{$\Phi_\text{sig} = \sum_{k=0}^{B} \Phi_\text{sig}[k]$ is the average signal photons per laser cycle.}. The number of photons detected by the SPAD pixel is stochastic and is governed by a Poisson process.
Over $N$ laser cycles, the number of photons detected in each time location $k$ is given by $\widehat{\Phi}[k] \sim \mathsf{Poisson}(N\Phi[k])$. 

%\clearpage
\section{Optimized Stepping Strategy for Proportional Binners\label{suppl:optimized_binner}}
This section presents additional background for the optimized proportional stepping binner defined in Eq.~(\ref{eq:gammadeltak}) in the main text, and also presents the process for tuning the various hyperparameters used in calculating the step size.

In general, the binner's control value is updated at each laser cycle according to $C_{n+1} = C_n + S_n$ where $C_n$ is the old CV, $S_n$ is the (signed) step size and $C_{n+1}$ is the new CV.
Recall that step size for a basic proportional binner (without any temporal decay or smoothing) tracking the $j^\text{th}$ boundary of $q$-quantile is computed as: 
\begin{equation} \label{eq:basicproportional}
\begin{aligned}
S_n = \left( \frac{j}{q} - \frac{E_n}{E_n + L_n}\right),
\end{aligned}
\end{equation}
where $E_n$ and $L_n$ denote the number of photons arriving earlier and later than the current CV. We refer to the right-hand side of Eq.~(\ref{eq:basicproportional}) by $\Delta_n$ in the further refinement of $S_n$ below. To speed up convergence and minimize variance once the binner converges, we modify the basic stepping strategy by introducing (i) a scaling factor ($K$), (ii) temporal decay ($\gamma$), and (iii) exponential smoothing factors $(\beta_1, \beta_2$) as described below.

\smallskip
\noindent
\textbf{Common experimental settings:} To find the best combination of optimal stepping parameters: $K$, $\gamma$,$\beta_1$, and $\beta_2$, we assume a Gaussian laser pulse of \SI{100}{\nano\second} repetition period (corresponding to a maximum distance range of \SI{15}{\m}) and full width at half maximum (FWHM) of \SI{0.32}{\nano\second}. 
We run each proportional binner for $L = 5000$ laser cycles and simulate the PEDH output with $q=32$ ED bins on a per-pixel basis.
We simulate 10 different distance values ranging from 1.5 meters to 13.5 meters, and for each distance, we consider eight (signal, background) photon-level pairs:
$(\Phi_\text{sig}, \Phi_\text{bkg}) \in \{$(1.0, 1.0), (1.0, 2.0), (1.0, 5.0), (1.0, 10.0),  (0.5, 0.5), (0.5, 1.0), (0.5, 2.5),  (0.5, 5.0) $\}.$ 
For comparison, we use two different metrics: the RMSE between the binner CV and the true quantile boundary it tracks (\textit{boundary RMSE}), and the \textit{distance RMSE} computed using the narrowest-bin-midpoint estimator $\widehat t_0$ for OEDH and PEDH.
To determine the best value for each optimized stepping parameter we simulate PEDH varying that parameter over a wide range of values with the rest of the parameters fixed, and finally compare the boundary RMSE and distance RMSE.

\subsection{Scaling Factor ($K$)}
\begin{figure}[ht]
\centering
\includegraphics[width=0.99\linewidth]{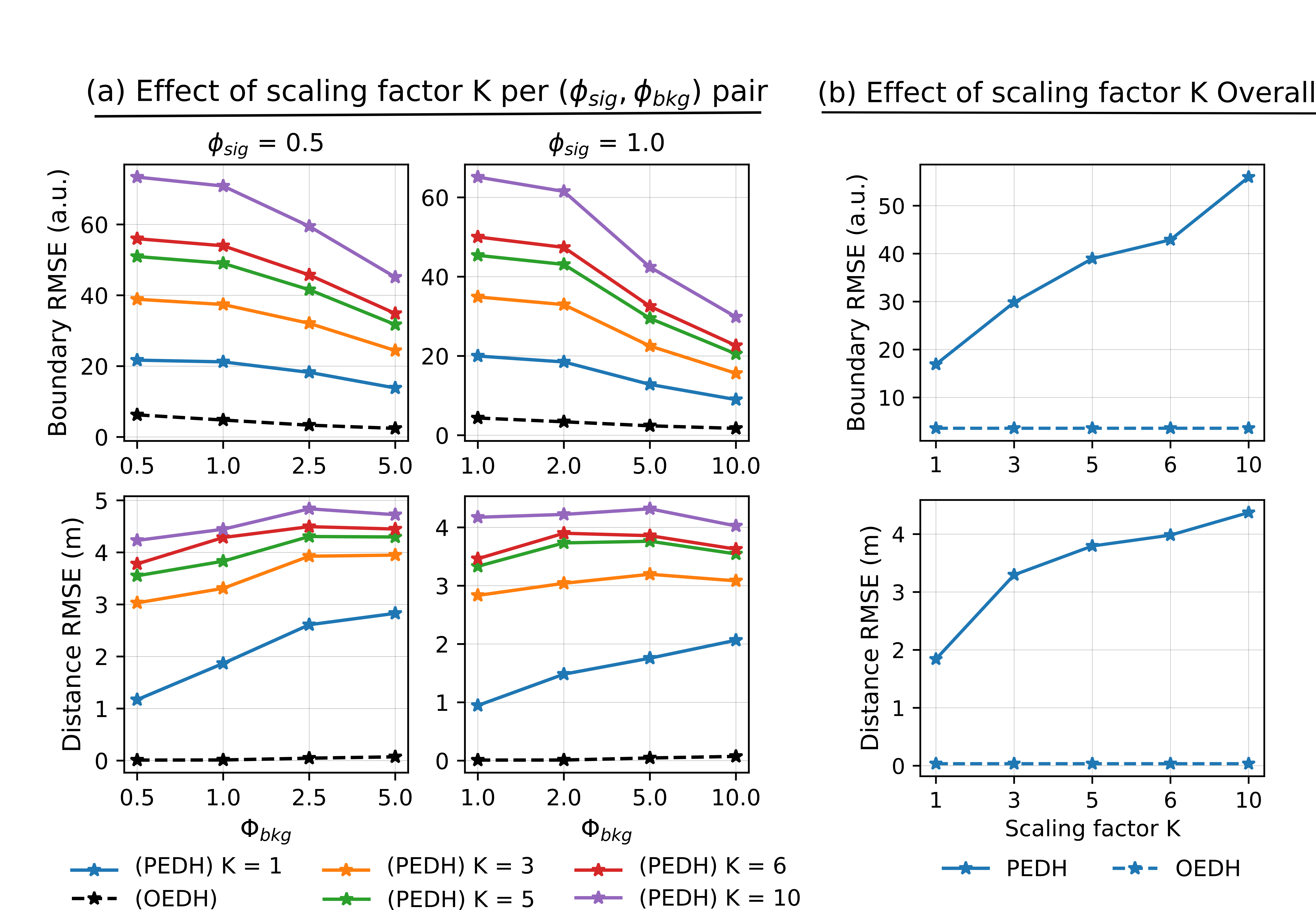}
\caption{\textbf{Choosing the best scaling factor $K$}: (a) Plots illustrate the effect of $K$ on different ($\Phi_\text{sig}, \Phi_\text{bkg}$) pairs for a wide range of distance values. (b) Plots illustrate the average errors over eight different ($\Phi_\text{sig}, \Phi_\text{bkg}$) pairs and varying scene distances. The average error increases as $K$ increases.}
\label{fig:Keffect}
\end{figure}

The basic proportional stepping strategy limits the step size $|S_n| \leq 1$ because the fractions $\frac{j}{q}$ and  $\frac{E_n}{(E_n + L_n)}$ can never be greater than $1$.
For faster convergence, we would like to allow larger step sizes that gradually become smaller as the CV approaches the true quantile location.
To allow larger step sizes, we multiply $\Delta_n$ by a scaling factor, which is a percentage of the number of time bins $B$. Hence for $K\%$, the step size with scaling factor is
\begin{equation} \label{eq:basicproportionalwithK}
\begin{aligned}
S_n = \frac{K}{100} B \Delta_n.
\end{aligned}
\end{equation}
A higher value of $K$ forces CV to be updated by larger step sizes, however, if $K$ is too high, then CV may cause large excursions around the true quantile location. 
We determine the best value of $K$ empirically by running simulations with fixed values of $\beta_1 = 0, \beta_2 = 0$, and $\gamma = 1$ and varying the value of $K$ and compare the boundary and distance RMSE. Suppl.~Fig.~\ref{fig:Keffect} demonstrates an increase in RMSE as $K$ is increased, which is evident as the increase in step size would results in high variance. Hence, if we do not apply any temporal decay (as $\gamma = 1$) or temporal smoothing (as $\beta_1 = 0, \beta_2 = 0$), choosing $K = 1$ would result in better performance than other values.  
%TODO[Done] Observe in Fig.~\ref{fig:Keffect} that XXXX hence we choose the value of K=??? for our simulations.

% \begin{figure}[h]
% \centering
% \includegraphics[width=0.8\linewidth]{supp_figs/Best_StepGain_Overall_init.png}
% \caption{\textbf{}: This plot illustrates the average errors over all the (signal, background) photon pairs, and scene distances. The average error increases as $K$ increases.}
% \label{fig:bestK}
% \end{figure}

\subsection{Temporal Decay Parameter ($\gamma$)}
In Eq.~(\ref{eq:gammadeltak}) we apply exponential smoothing by considering $\widetilde\Delta_n$: an exponentially decayed version of $\Delta_n$, which can be computed based on the decay parameter $\gamma$ as
\begin{equation} \label{eq:decayed}
\begin{aligned}
\widetilde\Delta_n =  \gamma^n\Delta_n.
\end{aligned}
\end{equation}
The choice of $\gamma$ is crucial: we would like $\gamma$ to be small enough that it limits the wandering around the true quantile location after convergence, but large enough that it does not prevent the CV from quickly approaching its final position.
We empirically determine the best value for gamma by simulating proportional binners with different values of $\gamma$, fixing $K = 1, \beta_1 = 0, \beta_2 = 0$.
%TODO[Done] Observe in Fig. \ref{fig:gamma}, hence we choose gamma = ???$

\begin{figure}[ht]
\centering
\includegraphics[width=0.99\linewidth]{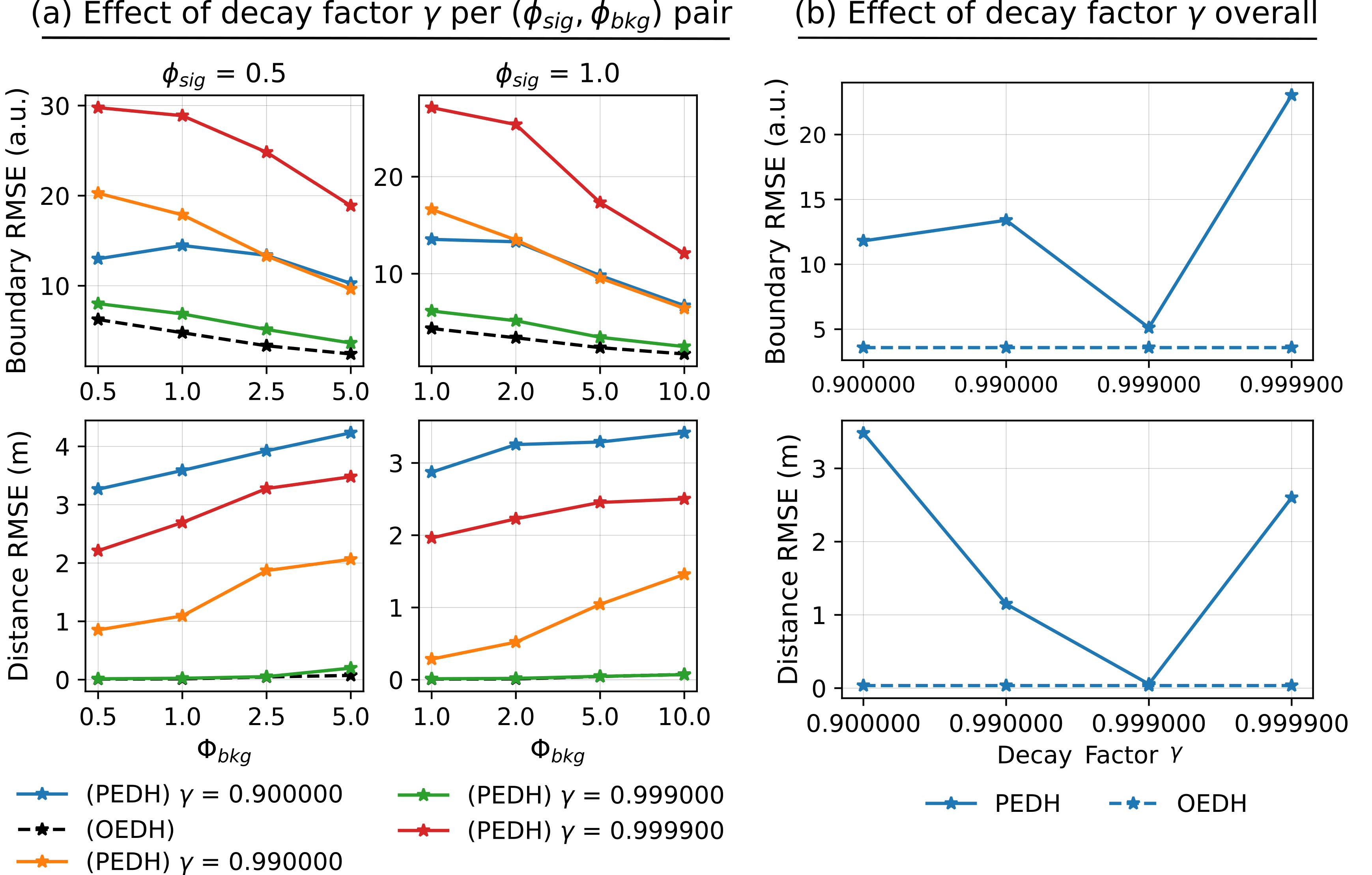}
\caption{\textbf{Choosing the best decay parameter $\gamma$}: (a) Plots illustrate the effect of $\gamma$ on different ($\Phi_\text{sig}, \Phi_\text{bkg}$) pairs for a range of distance values. (b) Plots illustrate the average errors over eight different ($\Phi_\text{sig}, \Phi_\text{bkg}$) pairs and varying scene distances. The average errors are least for $\gamma = 0.999$.}
\label{fig:gamma}
\end{figure}

% \begin{figure}[h]
% \centering
% \includegraphics[width=0.8\linewidth]{supp_figs/Best_DecayFactor_Overall_coarse.png}
% \caption{\textbf{Choosing the best decay parameter $\gamma$}: This plot illustrates the average errors over all the (signal, background) photon pairs, and scene distances for PEDH using different values of $\gamma$. The average errors are least for $\gamma = 0.999$.}
% \label{fig:bestK}
% \end{figure}
Observe that in Suppl.~Fig.~\ref{fig:gamma}, $\gamma = 0.999$ gives the lowest RMSE.
We further perform an exhaustive search over a narrower range of $\gamma$ values from $\gamma = 0.99901$ to $\gamma = 0.9999$ and find that $\gamma = 0.99902$ gives the best overall.
As an additional safeguard against extremely small step sizes, we do not reduce the decay factor beyond $n=4000$ laser cycles.

% we do not apply the decay factor beyond $n=4000$ laser cycles.[We do apply the decay factor but we don't reduce it any further]

\subsection{Exponential-Smoothing Parameters ($\beta_1, \beta_2$)}
We simulate proportional binners with different values of $\beta_2$ (fixing $\beta_1 = 0.5$). We then repeat the same experiment for different values of $\beta_1$ fixing $\beta_2$ to the empirical best value. 

\begin{figure}[ht]
\centering
\includegraphics[width=0.99\linewidth]{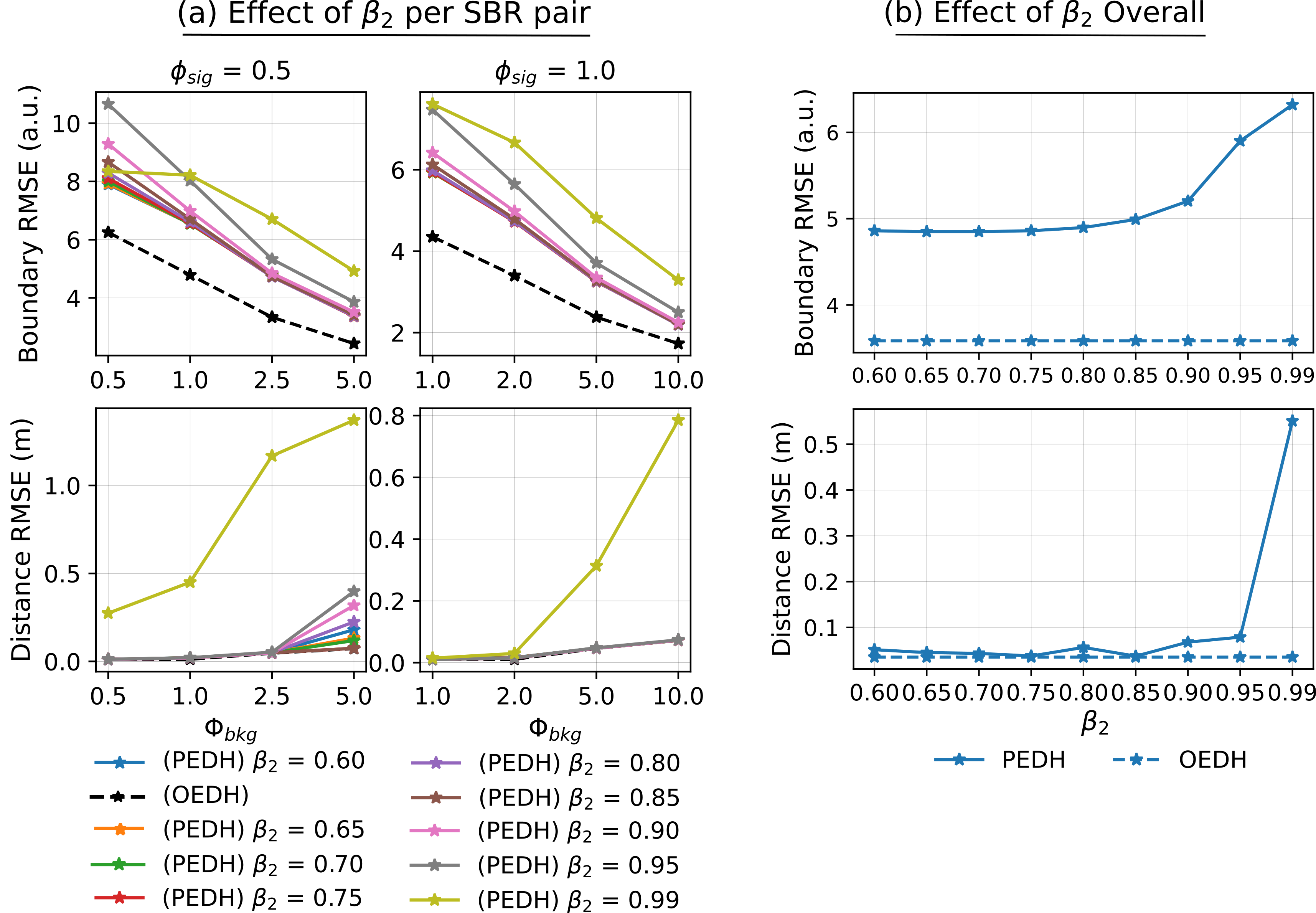}
\caption{\textbf{Choosing the best smoothing parameter $\beta_2$}: (a) Plots illustrate the effect of $\beta_2$ on different ($\Phi_\text{sig}, \Phi_\text{bkg}$) pairs for a range of distance values. (b) Plots illustrate the average errors over eight different ($\Phi_\text{sig}, \Phi_\text{bkg}$) pairs and varying scene distances. The average boundary RMSE starts increasing after $\beta_2 = 0.8$ and aver
age distance RMSE start increasing after $\beta_2 = 0.95$.}
\label{fig:beta2}
\end{figure}

\begin{figure}[ht]
\centering
\includegraphics[width=0.99\linewidth]{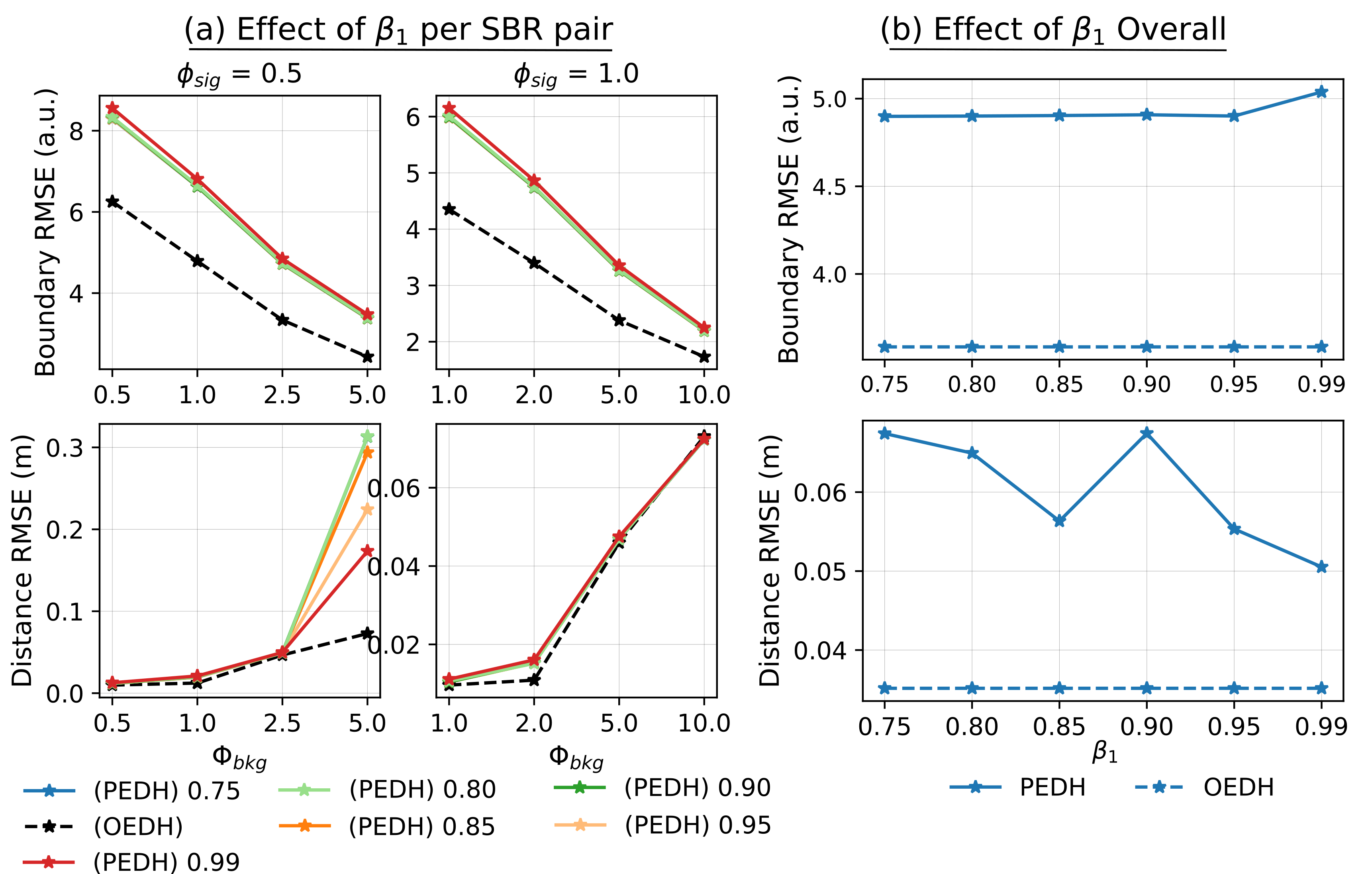}
\caption{\textbf{Choosing the best smoothing parameter $\beta_1$}: (a) Plots illustrate the effect of $\beta_1$ on different ($\Phi_\text{sig}, \Phi_\text{bkg}$) pairs for a range of distance values. (b) Plots illustrate the average errors over eight different ($\Phi_\text{sig}, \Phi_\text{bkg}$) pairs and varying scene distances. Considering boundary RMSE and distance RMSE $\beta_1 = 0.95$ is the best choice.}
\label{fig:beta1}
\end{figure}

In Suppl.~Fig.~\ref{fig:beta2} the average boundary RMSE starts increasing after $\beta_2 = 0.8$ and the average distance RMSE start increasing after $\beta_2 = 0.95$. As the overall value of boundary, RMSE is higher we choose $\beta_2 = 0.8$ as the best value for $\beta_2$. In Suppl.~Fig.~\ref{fig:beta1} the average boundary RMSE is almost the same for all values of $\beta_1$ except 0.99 and the distance RMSE is lower for $\beta_1 = $ 0.85, 0.95, and 0.99, hence overall we choose $\beta_1 = 0.95$.

%TODO[Done] Observe in Fig.~\ref{fig:beta1} and \ref{fig:beta2} ??????

Upon choosing $\gamma = 0.99902, \beta_1 = 0.95, \beta_2 = 0.8$ we observe that error values are close for $K=1$ and $K=3$ hence we choose $K=3$ for faster convergence. 

% \smallskip
% \noindent
% \textbf{Common experimental settings:} To find the best combination of optimal stepping parameters we assume a Gaussian laser pulse of \SI{100}{\nano\second} repetition period (corresponding to a maximum distance range of \SI{15}{\m}) and full width at half maximum (FWHM) of \SI{0.32}{\nano\second}. 
% We run each proportional binner for $L = 5000$ laser cycles and simulate the PEDH output with $q=32$ ED bins on a per-pixel basis.
% We simulate 10 different distance values ranging from 1.5 meters to 13.5 meters, and for each distance, we consider eight (signal, background) photon-level pairs:
% $(\Phi_\text{sig}, \Phi_\text{bkg}) \in \{$(1.0, 1.0), (1.0, 2.0), (1.0, 5.0), (1.0, 10.0),  (0.5, 0.5), (0.5, 1.0), (0.5, 2.5),  (0.5, 5.0) $\}.$ 
% For comparison, we use two different metrics: the RMSE between the binner CV and the true quantile boundary it tracks (\textit{boundary RMSE}), and the RMSE in the distance estimates computed using the narrowest-bin-midpoint estimator $\widehat t_0$ for OEDH and EDH obtained from proportional binners \textit{distance RMSE}. 
% To determine the best value for each optimized stepping parameter we simulate PEDH varying that parameter over a wide range of values with the rest of the parameters fixed, and finally compare the boundary RMSE and distance RMSE.

\subsection{Fixed-Stepping vs. Optimized Proportional-Stepping}
% We provide evaluation results to highlight the improvement we achieve in terms of binner convergence with our optimized stepping strategy as compared to the basic strategy (constant step size). 
% We consider the same experimental settings and simulate PEDH using basic proportional stepping (with $K=1, \beta_1 = 0, \beta_2 = 0, \gamma = 1$), and optimized proportional stepping (with $K=3, \beta_1 = 0.95, \beta_2 = 0.8, \gamma = 0.99902$) and compare the boundary RMSE and distance RMSE. 
% The results in Suppl.~Fig.~\ref{fig:stepcompare} highlights that by using the optimized-stepping proportional binner we can obtain quantile boundary estimates that are significantly closer to the oracle values as compared to the the estimates obtained from the basic proportional binner as all the errors are significantly lower when we use the optimized stepping strategy. 

% \begin{figure}[ht]
% \centering
% \includegraphics[width=0.6\linewidth]{supp_figs/SimplevsOptimised.png}
% \caption{\textbf{Optimized stepping performs better than basic stepping}: This plot compares the performance of a basic proportional stepping binner and the optimized proportional stepping binner for a range of distance values and different (signal, background) photon-level pairs. The optimized stepping binner can achieve significantly better performance than the basic stepping binner and its performance is close to the oracle values.}
% \label{fig:stepcompare}
% \end{figure}

We simulated median tracking binners with the fixed-stepping strategy and our proposed optimized proportional stepping strategy for 10 scene distances and 6 illumination conditions over 500 Monte Carlo runs. 
As all the binners run for the same exposure of 5000 laser cycles, the lower RMSE between the true median and the binner control values (CVs) shared in Table \ref{table:median-hedhvpedh-t1} demonstrate that our optimized stepping binner converges faster and closer to the true median and is more robust to changing illumination conditions than the fixed-stepping binner used by Ingle and Maier \cite{racingspad}.

\begin{table}[t]
\caption{{\small Quantitative metrics for simulated median tracking binners show large improvements in RMSE, faster convergence, and robustness to illumination conditions for optimized stepping.}}
\vspace{-0.1in}
\centering
\setlength{\tabcolsep}{10pt}
\label{table:median-hedhvpedh-t1}
{\begin{tabular}{lllll}
\hline
Background flux ($\Phi_\text{bkg}$)                                                      & 0.5 & 1.0 & 2.0 & 5.0  \\ \hline
\begin{tabular}[c]{@{}c@{}}Fixed step \cite{racingspad}\end{tabular}      & 6.50 & 7.32 & 9.39 & 10.13  \\
\begin{tabular}[c]{@{}c@{}}Optimized step [Ours]\end{tabular}  & 5.46 & 4.20 & 4.95 & 4.06   \\ \hline
\end{tabular}}
\vspace{-0.1in}
\end{table}

%TODO[Done]: Observe in fig:stepcompare ...???

\subsection{Theoretical Proof of Convergence}
The Markov chain model presented by Ingle \& Maier \cite{racingspad} can be modified to handle proportional stepping and derive stationary distributions to quantify convergence.
It is difficult to analyze the theoretical convergence properties of the optimized binner due to the different temporal smoothing terms.
Here we analyze the convergence of the basic proportional binner that decides its step size based on the ratio of early (E) to late (L) photons.
In future theoretical work we will explore connections with control theory (PID control) and gradient descent to analyze convergence rates of the optimized binner.
A proof sketch for the $25^\textbf{th}$ percentile ($Q_{25}$) proportional binner is as follows.
Without loss of generality, suppose the binner's $\text{CV} \!> \!Q_{25}.$
The probability that $E\! >\! 3L$ is exponentially small (bounded using a Chernoff bound).
Therefore, the CV will move towards $Q_{25}$ with high probability.

\subsection{Single-Pixel Distance Estimators}

Recall that for an ED histogram that tracks arbitrary $q$-quantiles, with bin boundary locations given by $\{ t_i \}_{i=0}^q$, we define the piecewise constant local photon density estimator as $\rho_0(t) = \nicefrac{1}{(t_j - t_{j-1})}$ for $t_{j-1} \leq t < t_j$, $1\leq j \leq q$ and $0 \leq t \leq B$. And we also propose a piecewise linear local photon density estimate $\rho_1(t)$ which is obtained by taking the sequence of non-uniformly spaced pairs of points $\left\{(\nicefrac{(t_{j}-t_{j-1})}{2}, \nicefrac{1}{(t_j-t_{j-1})} \right\}_{j=1}^q$ interpolated on a grid of $1024$ uniformly spaced discrete time locations between $0$ and $B$ as follows 
\begin{equation} \label{eq:rho1} 
\begin{aligned}
  \rho_1(t) = \left( \frac{\rho_0(t_j) - \rho_0(t_{j-1})}{t_j - t_{j-1}}\right)( t - t_j) + \rho_0(t_j).
\end{aligned}
\end{equation}
We also define an alternative time-of-flight estimate as $\widehat t_1 = \frac{1}{2} \argmax_{t \in [0,B]} \rho_1(t).$ 

\begin{figure}[ht]
\centering
\includegraphics[width=0.95\linewidth]{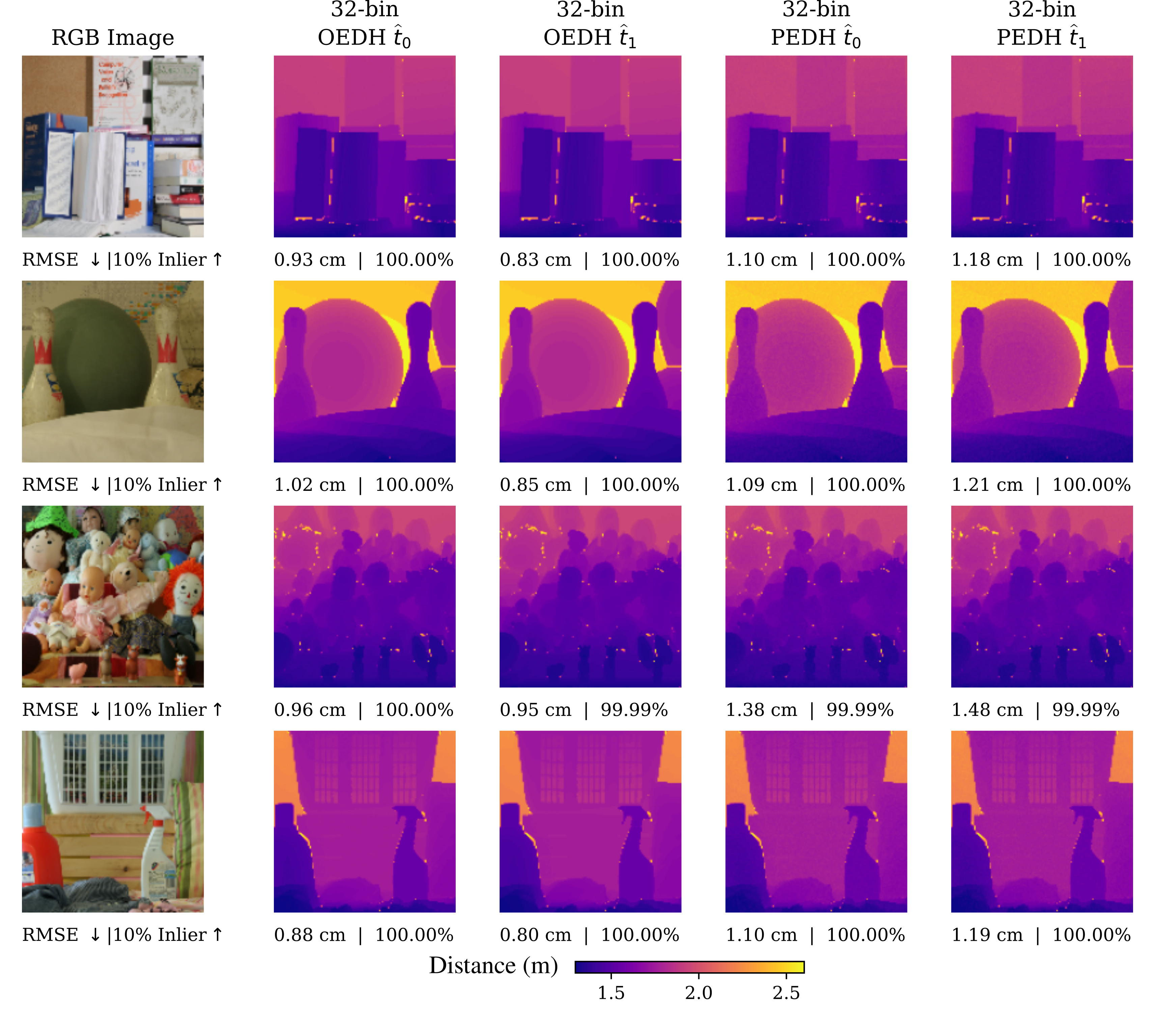}
\caption{\textbf{Performance comparison of $\widehat t_0$ and $\widehat t_1$} estimators: Quantitative and qualitative results for distance estimates obtained from narrowest-bin-midpoint distance estimator $\widehat t_0$ and linearly interpolated distance estimator $\widehat t_1$ for OEDH and PEHD. $\widehat t_1$ improves distance estimates over the narrowest-bin method for OEDH
but deteriorates the distance estimates for PEDH.}
\label{fig:rho0rho1qual}
\end{figure}

Figure~\ref{fig:rho0rho1} in the main text shows an example of the effect of linear interpolation on photon density estimates obtained from OEDH and PEDH and highlight how the noise in PEDH boundaries (although small) results in strong noise near the peak of $\rho_0$ and shifts the peak of $\rho_1$ away from the true peak.
To further support this observation we provide $\widehat t_0$ and $\widehat t_1$ distance estimates for multiple scenes in Suppl.~Fig.~\ref{fig:rho0rho1qual}.
The results further support the observation that $\widehat t_1$ improves distance estimates over $\widehat t_0$ for OEDH but deteriorates the distance estimates for PEDH.
To denoise $\rho_1$ obtained from PEDH and to additionally exploit correlations across neighboring pixel measurements we develop a data-driven approach and train DNN to take the noisy $\rho_1$ estimates from PEDH as input and exploit the spatiotemporal correlations to predict high-quality distance maps.
Since $\rho_1$ results in better distance estimates than $\rho_0$ for the oracle EDH, we pass $\rho_1$ as the input to the DNN instead of $\rho_0$.

%-----------------------------------------------------
\clearpage
\section{Experimental Results Comparing HEDH and PEDH\label{suppl:hedh_vs_pedh}}
In this section, we show additional experimental results comparing the performance of HEDH and PEDH. We show binner-trajectory plots for 8-bin HEDH and 8-bin PEDH for different illumination conditions and compare the distance estimates obtained by running the histogrammers for a real hardware dataset \cite{Gupta2019AsynchronousS3} and a flash LiDAR dataset \cite{itof2dtof}. 

\subsection{Comparing CV Update Trajectories}
\begin{figure*}[ht]
\centering
\includegraphics[width=0.99\linewidth]{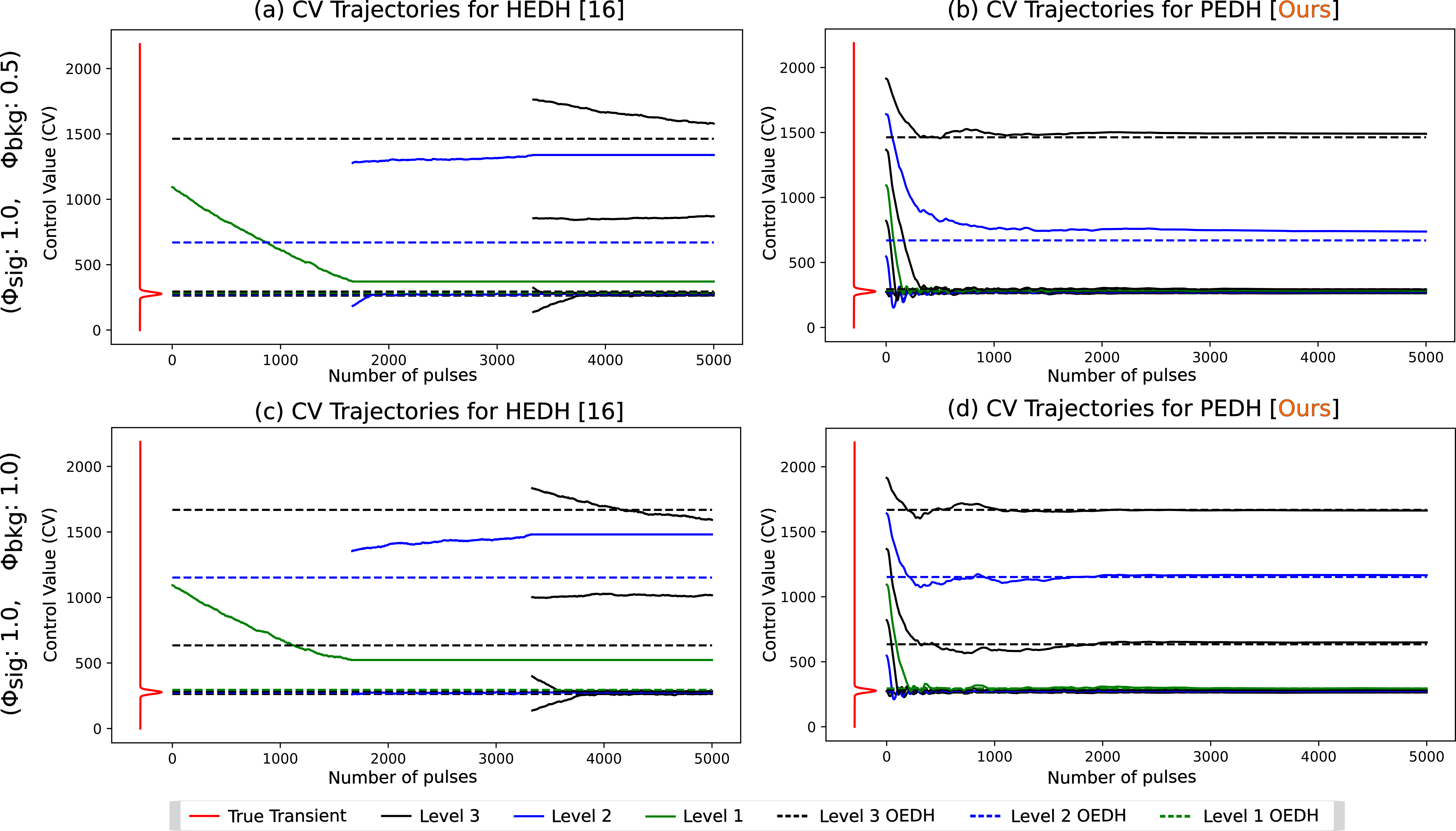}
\caption{\textbf{Improved binner trajectories}: Binner-trajectory plots for 8-bin HEDH and 8-bin PEDH demonstrate that binners of PEDH track different quantiles more accurately, leveraging the complete exposure time as compared to the binners of HEDH.}
\label{suppl:cvtraj_hedh_vs_pedh}
\end{figure*}

We simulated 8-bin HEDH and 8-bin PEDH for different illumination conditions to compare the binner trajectories of both histogrammers. 
Trajectory plots in Suppl.~Fig.~\ref{suppl:cvtraj_hedh_vs_pedh} demonstrate that the PEDH binners achieve better convergence and lower variance.
We also observe that as the background photons increase binner estimates are severely affected in the case of HEDH but not in the case of PEDH, highlighting the robustness of PEDH to different illumination conditions. 
These observations are consistent with the results shown by Ingle and Maier \cite{racingspad} where HEDH performs well for high signal and low background photon conditions.

\subsection{Emulation Results for Real Hardware Data}
We conducted hardware emulation experiments on a publicly available hardware dataset by Gupta et al. \cite{Gupta2019AsynchronousS3} which contains real-world noise sources (dark counts, afterpulsing, dead-time).
Initial hardware emulation results shown by Ingle and Maier use an HEDH with a fixed-stepping strategy in a single illumination condition $(\Phi_\text{sig}, \Phi_\text{bkg}) = (2.0,0.2)$.
We emulated 8-bin HEDH and 8-bin PEDH for 5000 laser cycles for 5 different illumination conditions.
For the same parameters, we also simulated HEDH and PEDH for a flash LiDAR dataset by Barragan et al. \cite{itof2dtof} (``iToF2dToF'' dataset).

\begin{figure*}[t]
\centering
\includegraphics[width=0.99\linewidth]{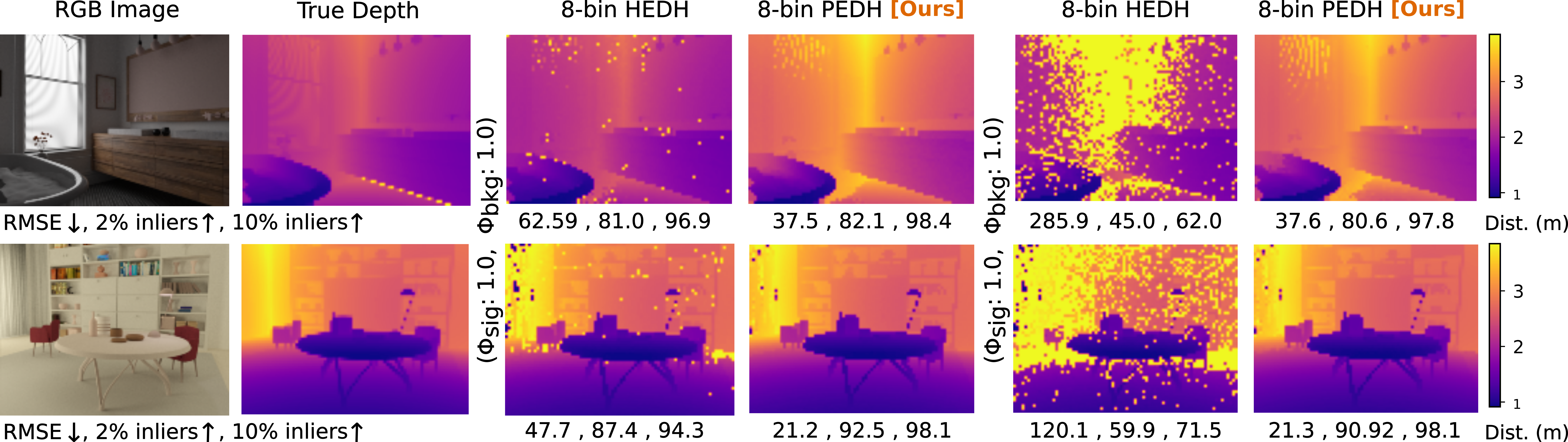}
\caption{\textbf{Evaluation results on ``iToF2dToF'' dataset \cite{itof2dtof}}: Qualitative and quantitative results for (signal, background) photon pairs of (1.0, 1.0) and (1.0, 2.0) demonstrate that distance estimates obtained from PEDH are more accurate and robust to changing illumination conditions.}
\label{suppl:itof2dtof_results}
\end{figure*}

Table~\ref{suppl:table_hardware-results} contains average RMSE error and 2\% inliers over different scenes for 5 different illumination conditions and Suppl.~Fig.~\ref{suppl:itof2dtof_results} contains qualitative and quantitative results for the ``iToF2dToF'' dataset.

\begin{table}[hb]
\centering
\setlength{\tabcolsep}{5pt}
\caption{Metrics for the real hardware dataset by Gupta et al.\cite{Gupta2019AsynchronousS3} and FlashLiDAR dataset by Barragan et al. show large improvements in RMSE and 2\% inliers, and robustness to lighting conditions with our PEDH over HEDH \cite{racingspad}}
\label{suppl:table_hardware-results}
\vspace{-0.1in}
\begin{tabular}{c|cccc|cccc}
\hline
\multicolumn{1}{l|}{} & \multicolumn{4}{c|}{Hardware Dataset \cite{Gupta2019AsynchronousS3}}                                                                                                                                                                                                                                                               & \multicolumn{4}{c}{``iToF2dToF'' Dataset\cite{itof2dtof}}                                                                                                                                                                                                                                                                    \\ \hline
                      & \multicolumn{2}{c|}{RMSE (cm)}                                                                                                                                & \multicolumn{2}{c|}{2 \% inlier}                                                                                                        & \multicolumn{2}{c|}{RMSE (cm)}                                                                                                                                & \multicolumn{2}{c}{2 \% inlier}                                                                                                         \\ \hline
$\Phi_\text{bkg}$               & \multicolumn{1}{c|}{\begin{tabular}[c]{@{}c@{}}HEDH\end{tabular}} & \multicolumn{1}{c|}{\begin{tabular}[c]{@{}c@{}}PEDH\end{tabular}} & \multicolumn{1}{c|}{\begin{tabular}[c]{@{}c@{}}HEDH\end{tabular}} & \begin{tabular}[c]{@{}c@{}}PEDH\end{tabular} & \multicolumn{1}{c|}{\begin{tabular}[c]{@{}c@{}}HEDH\end{tabular}} & \multicolumn{1}{c|}{\begin{tabular}[c]{@{}c@{}}PEDH\end{tabular}} & \multicolumn{1}{c|}{\begin{tabular}[c]{@{}c@{}}HEDH\end{tabular}} & \begin{tabular}[c]{@{}c@{}}PEDH\end{tabular} \\ \hline
0.1                   & \multicolumn{1}{c|}{16.6}                                                   & \multicolumn{1}{c|}{16.2}                                                     & \multicolumn{1}{c|}{92.6}                                                  & 91.8                                                    & \multicolumn{1}{c|}{38.9}                                                   & \multicolumn{1}{c|}{37.4}                                                     & \multicolumn{1}{c|}{88.6}                                                  & 91.6                                                    \\
0.5                   & \multicolumn{1}{c|}{17.3}                                                   & \multicolumn{1}{c|}{15.9}                                                     & \multicolumn{1}{c|}{89.1}                                                  & 90.1                                                    & \multicolumn{1}{c|}{46.8}                                                   & \multicolumn{1}{c|}{38.6}                                                     & \multicolumn{1}{c|}{86.7}                                                  & 91.3                                                    \\
1.0                   & \multicolumn{1}{c|}{22.6}                                                   & \multicolumn{1}{c|}{16.4}                                                     & \multicolumn{1}{c|}{77.8}                                                  & 77.2                                                   & \multicolumn{1}{c|}{56.4}                                                   & \multicolumn{1}{c|}{38.2}                                                     & \multicolumn{1}{c|}{79.6}                                                  & 90.6                                                    \\
2.0                   & \multicolumn{1}{c|}{78.6}                                                   & \multicolumn{1}{c|}{16.2}                                                     & \multicolumn{1}{c|}{43.4}                                                  & 54.2                                                    & \multicolumn{1}{c|}{143.2}                                                   & \multicolumn{1}{c|}{39.1}                                                     & \multicolumn{1}{c|}{46.2}                                                  & 83.2                                                    \\
5.0                   & \multicolumn{1}{c|}{153.8}                                                   & \multicolumn{1}{c|}{17.2}                                                     & \multicolumn{1}{c|}{0.3}                                                   & 37.2                                                    & \multicolumn{1}{c|}{230.3}                                                   & \multicolumn{1}{c|}{45.9}                                                     & \multicolumn{1}{c|}{6.9}                                                   & 34.2                                                    \\ \hline
% 10.0                  & \multicolumn{1}{c|}{66.0}                                                   & \multicolumn{1}{c|}{18.4}                                                     & \multicolumn{1}{c|}{0.3}                                                   & 11.3                                                    & \multicolumn{1}{c|}{220.2}                                                   & \multicolumn{1}{c|}{52.1}                                                     & \multicolumn{1}{c|}{3.1}                                                   & 15.9                                                    \\ \hline
\end{tabular}
\end{table}

%-----------------------------------------------------

\clearpage
\section{Additional Distance Estimation Results \label{suppl:distance_estimation}}

\begin{figure*}[ht]
\centering
\includegraphics[width=0.99\linewidth]{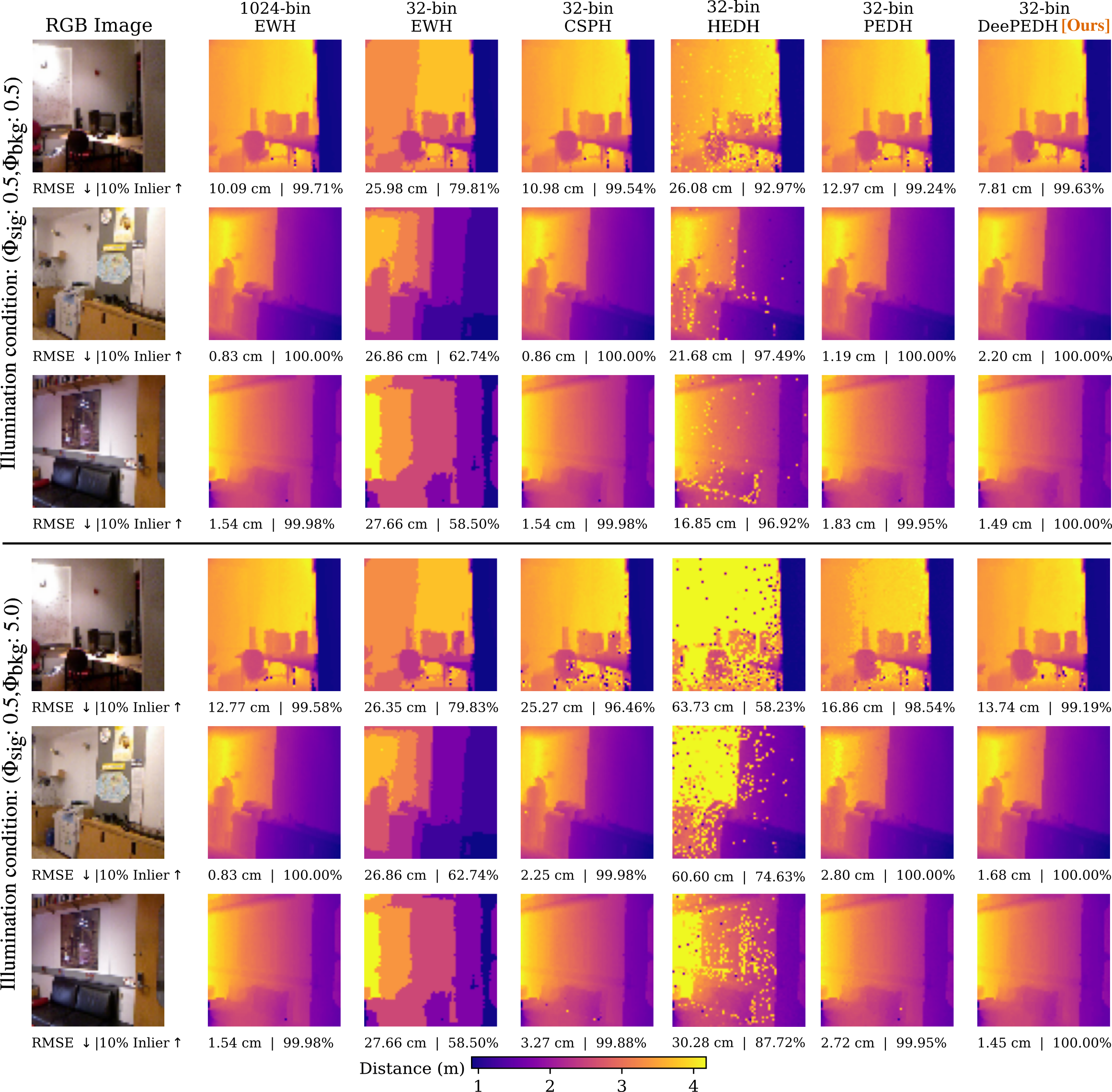}
\caption{\textbf{Evaluation results on NYUv2 test images}: Qualitative and quantitative results for (signal, background) photon pairs of (0.5, 0.5) and (0.5, 5.0). We observe that the $\widehat t_0$ PEDH estimator generates noisy distance estimates for scene objects that are darker and/or are at a farther distance from the camera, whereas DeePEDH can handle such scenarios and provide better distance estimates.}
\label{fig:distestim_results_sbr1}
\end{figure*}

\begin{figure*}[ht]
\centering
\includegraphics[width=0.99\linewidth]{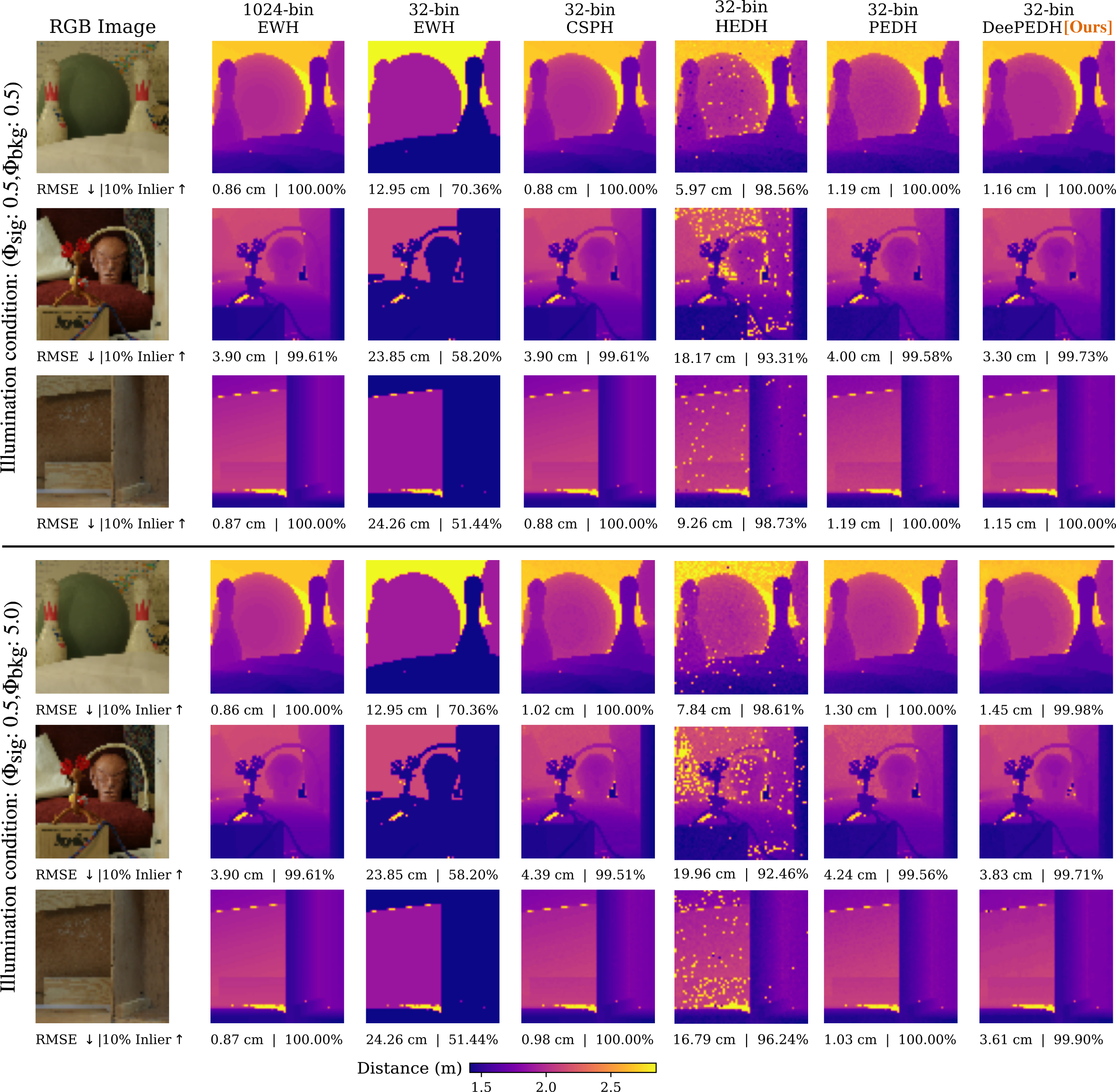}
\caption{\textbf{Evaluation results on Middlebury test images}: Qualitative and quantitative results for (signal, background) photon pairs of (0.5, 0.5) and (0.5, 5.0). We observe that DeePEDH can generalize well on the Middlebury dataset, which is much different from the training dataset in terms of distance ranges, lighting conditions, and scene objects.}
\label{fig:distestim_results_sbr2}
\end{figure*}

In this section, we show additional qualitative and quantitative results for distance estimation using DeePEDH.
For comparison, we show the results for the narrowest-bin-midpoint estimator $\widehat t_0$ and interpolated estimator $\widehat t_1$ along with other baselines. 
Suppl.~Fig.~\ref{fig:distestim_results_sbr1} and Suppl.~Table~\ref{table:quant-supp-table1} 
 contain qualitative and quantitative distance estimation results for NYUv2 test images and Suppl.~Fig.~\ref{fig:distestim_results_sbr2} and Suppl.~Table~\ref{table:quant-supp-table2} 
 contain qualitative and quantitative distance estimation results for Middlebury test images.

\begin{table}[ht]
\caption{Quantitative distance estimation results comparing the average performance of conventional (EWH-based) SPCs, CSPH\cite{csph}, narrowest bin PEDH distance estimator ($\widehat t_0$), linear interpolation based PEDH distance estimator $\widehat t_1$ and DeePEDH for 10 NYUv2 test set images and 8 different (signal, background) photon-level pairs.}
\label{table:quant-supp-table1}
\vspace{-0.1in}
\begin{tabular}{r|llllll}
\multicolumn{1}{c|}{\begin{tabular}[c]{@{}c@{}}NYUv2 \\ Dataset\end{tabular}} & \begin{tabular}[c]{@{}l@{}}Conventional\\ (1024-bin)\end{tabular} & \begin{tabular}[c]{@{}l@{}}Conventional\\ (32-bin)\end{tabular} & \begin{tabular}[c]{@{}l@{}}CSPH {\cite{csph}}\\ (32-bin)\end{tabular} & \begin{tabular}[c]{@{}l@{}}PEDH $\widehat t_0$\\ (32-bin)\end{tabular} & \begin{tabular}[c]{@{}l@{}}PEDH $\widehat t_1$\\ (32-bin)\end{tabular} & \begin{tabular}[c]{@{}l@{}}DeePEDH \\ \textcolor{RedOrange}{[Ours]} (32 bin)\end{tabular} \\ \hline
RMSE (cm) & 6.89 & 29.80 & 15.01 & 18.05 & 18.22 & 10.92 \\
MAE (cm) & 1.06 & 24.58 & 2.03 & 2.40 & 2.77 & 1.84 \\
2\% Inliers & 99.89 & 10.69 & 99.45 & 97.87 & 95.06 & 98.51 \\
10\% Inliers & 99.90 & 55.86 & 99.62 & 99.71 & 99.71 & 99.81 \\ \hline
\end{tabular}
\end{table}

\begin{table}[ht]
\caption{Quantitative distance estimation results comparing the average performance of conventional (EWH-based) SPCs, CSPH\cite{csph}, narrowest bin PEDH distance estimator ($\widehat t_0$), linear interpolation based PEDH distance estimator $\widehat t_1$ and DeePEDH for 10 Middlebury test set images and 8 different (signal, background) photon-level pairs.}
\label{table:quant-supp-table2}
\vspace{-0.1in}
\begin{tabular}{r|llllll}
\multicolumn{1}{c|}{\begin{tabular}[c]{@{}c@{}}Middlebury \\ Dataset\end{tabular}} & \begin{tabular}[c]{@{}l@{}}Conventional\\ (1024-bin)\end{tabular} & \begin{tabular}[c]{@{}l@{}}Conventional\\ (32-bin)\end{tabular} & \begin{tabular}[c]{@{}l@{}}CSPH {\cite{csph}}\\ (32-bin)\end{tabular} & \begin{tabular}[c]{@{}l@{}}PEDH $\widehat t_0$\\ (32-bin)\end{tabular} & \begin{tabular}[c]{@{}l@{}}PEDH $\widehat t_1$\\ (32-bin)\end{tabular} & \begin{tabular}[c]{@{}l@{}}DeePEDH \\ \textcolor{RedOrange}{[Ours]} (32 bin)\end{tabular} \\ \hline
RMSE (cm) & 2.00 & 26.93 & 2.79 & 2.47 & 2.99 & 2.56 \\
MAE (cm) & 0.76 & 23.47 & 0.84 & 0.91 & 1.32 & 0.99 \\
2\% Inliers & 99.96 & 4.54 & 99.93 & 99.64 & 94.83 & 99.13 \\
10\% Inliers & 99.96 & 35.46 & 99.95 & 99.96 & 99.96 & 99.92 \\ \hline
\end{tabular}
\end{table}

\clearpage
\section{Additional Results on Visual Odometry and TSDF Fusion \label{suppl:voAndtsdf}}
\begin{figure}[ht]
\centering
\includegraphics[width=0.95\linewidth]{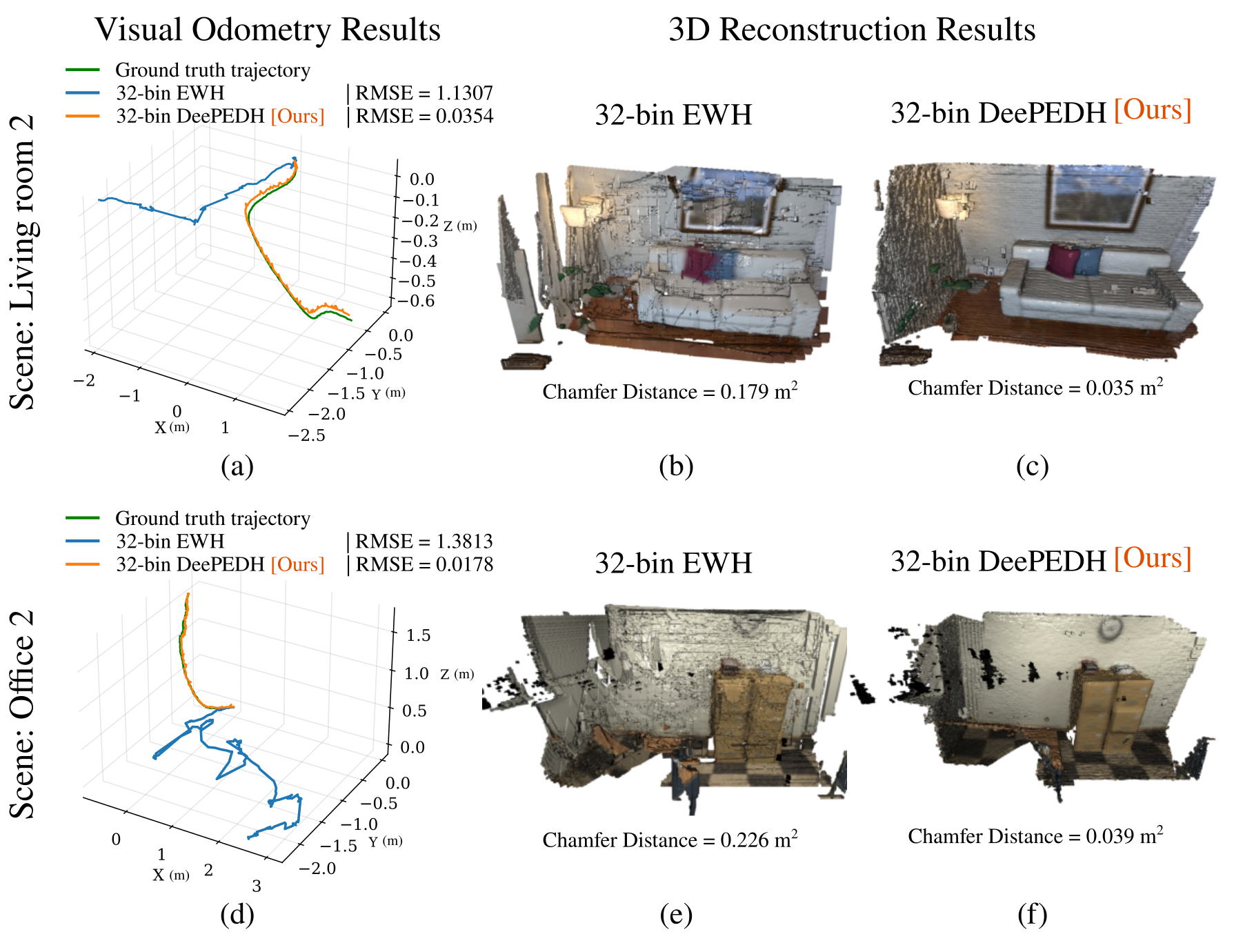}
\caption{\textbf{Visual odometry and 3D reconstruction results.} DeePEDH provides better camera tracking and preserves fine scene structures.}
\label{fig:voresults_2}
\end{figure}

% \begin{figure}[h]
% \centering
% \includegraphics[width=0.99\linewidth]{supp_figs/VO_Livingroom1.png}
% \caption{\textbf{Visual odometry and 3D reconstruction results for \textit{Living room 1} sequence.}}
% \label{fig:voresults_3}
% \end{figure}

% \begin{figure}[h]
% \centering
% \includegraphics[width=0.99\linewidth]{supp_figs/VO_Livingroom2.png}
% \caption{\textbf{Visual odometry and 3D reconstruction results for \textit{Living room 2} sequence.}}
% \label{fig:voresults_4}
% \end{figure}

The results in Suppl.~Fig.~\ref{fig:voresults_2} demonstrate that DeePEDH SPCs enable better camera tracking and 3D reconstruction.
The camera motion trajectory reconstructed from DeePEDH distance maps (green) provides over $10\times$ lower RMSE than the trajectory estimated using a 32-bin EWH. 
Using DeePEDH SPCs results in high quality 3D reconstructions both qualitatively and in terms of quantitative metrics.

%-----------------------------------------------------

\clearpage
\section{Additional Results on Semantic Segmentation \label{suppl:rgbdseg}}

\begin{figure}[ht]
\centering
\includegraphics[width=0.8\linewidth]{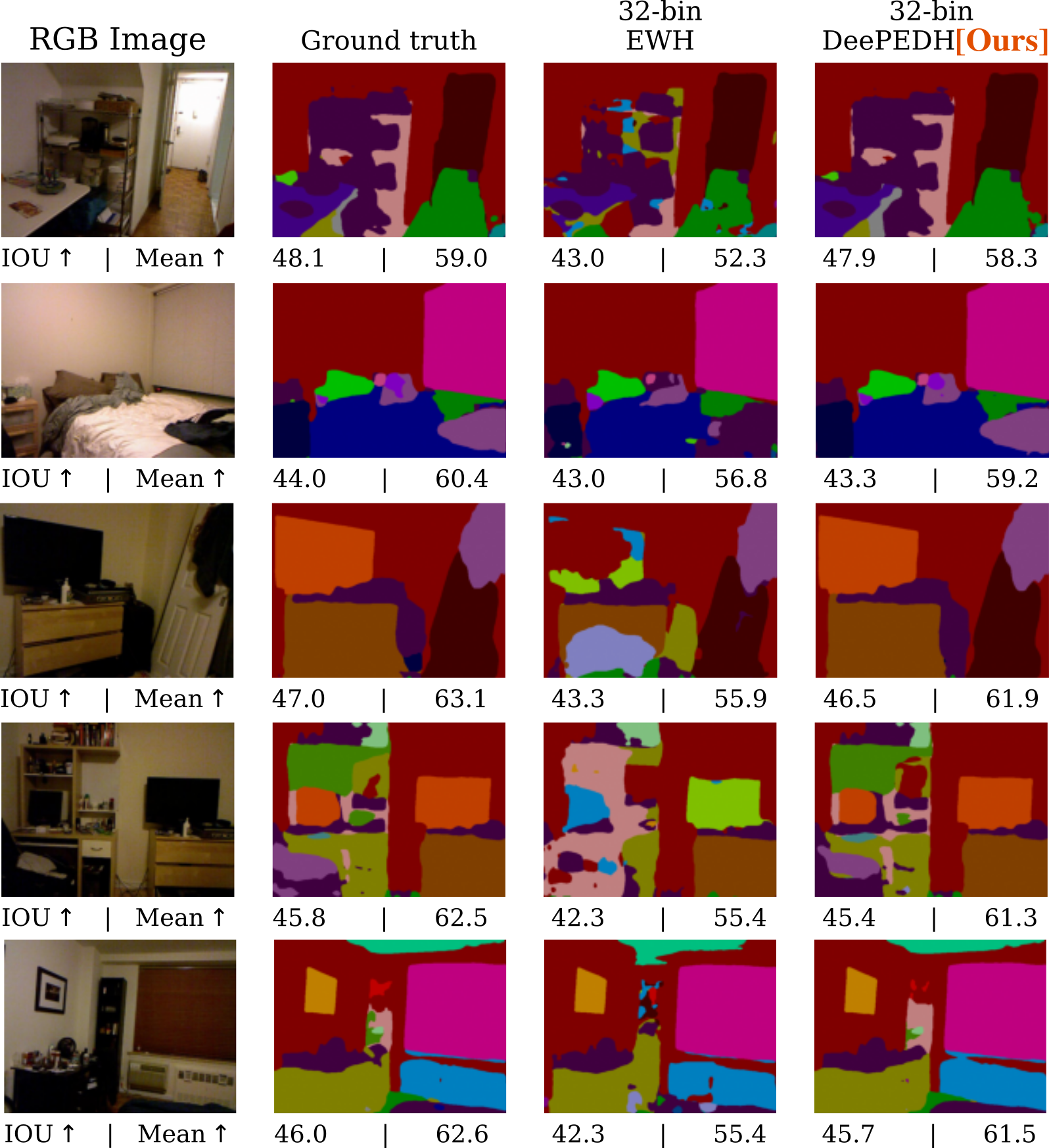}
\caption{\textbf{PEDH SPCs improve performance of deep RGBD semantic segmentation models}: Figure showing semantic segmentation results for a pre-trained CEN network \cite{cen} using distance estimates from 32-bin EWH and 32-bin DeePEDH.
As the distance maps from DeePEDH are significantly closer to the ground truth, the segmentation results are better than using distance images from 32-bin EWH. The effect of good distance estimates is evident for regions with low albedo where the RGB image fails to provide any reliable features and the model needs to depend on accurate scene distance estimates for useful features.}
\label{fig:rgbdseg_supp}
\end{figure}

The results in Suppl.~Fig.~\ref{fig:rgbdseg_supp} demonstrate that DeePEDH SPCs improve the performance of RGBD semantic segmentation models especially for regions where RGB images contain less texture information.

\end{document}